\documentclass[prl,twocolumn, amsmath,amssymb, aps,showkeys,]{revtex4-2}

\usepackage{CJK} 

\usepackage{graphicx}
\usepackage{dcolumn}
\usepackage{bm}
\usepackage{graphicx} 

\usepackage[allcolors=black]{hyperref}
\usepackage[capitalise]{cleveref}
\usepackage{xcolor} 

\usepackage{ulem}

\newcommand\lya{{Ly{$\alpha$}}}

\date{\today}

\begin{document}

\title{Tide-and-Seek: Anisotropic scale-dependent bias in the Lyman-alpha forest and designer samples}
\author{Yong Sheng Yap \begin{CJK*}{UTF8}{bsmi}(葉勇陞)\end{CJK*} }\affiliation{ DAMTP, Centre for Mathematical Sciences, University of Cambridge, Wilberforce Road, Cambridge CB3 0WA, UK.}
\affiliation{Kavli Institute for Cosmology Cambridge, Madingley Road, Cambridge CB3 0HA, UK.}
\author{William R. Coulton }\affiliation{Department of Physics, University of Oxford, Denys Wilkinson Building, Keble Road, Oxford OX1 3RH, UK.}
\author{Vid Ir\v{s}i\v{c} }
\affiliation{Centre for Astrophysics Research, Department of Physics, Astronomy and Mathematics, University of Hertfordshire, College Lane, Hatfield, AL10 9AB, UK}
\begin{abstract}
    Searches for primordial non-Gaussianity are powerful ways of learning new information about the early Universe. Observing primordial signals in the late Universe is challenging as, in general, it is difficult to disentangle the small, primordial signatures from the complex physics that describes the distribution of galaxies and large-scale cosmic structures. We demonstrate that scale-dependent bias from primordial non-Gaussianity with non-trivial directional dependence in the squeezed limit (also known as anisotropic primordial non-Gaussianity) can arise in a much larger class of tracers than previously considered. This is a smoking gun signature of primordial Universe physics that cannot be mimicked by late-Universe processes. We demonstrate this explicitly with simulations of the Lyman-$\alpha$ forest and halos selected through their projected shape. The former is a standard cosmological probe and standard analyses could be simply extended to constrain this. The latter provides an interesting new approach: design samples with strong selection effects to provide new sensitivity to primordial physics. As a demonstration of these concepts, we apply this method to BOSS DR12 data, finding no evidence of anisotropic non-Gaussianity.
    \end{abstract}
\keywords{Large scale structure of the Universe, Inflation}
\maketitle

Over the past few decades, cosmological observations have painted a remarkably simple picture of the primordial Universe, in which the initial fluctuations that seeded structure today are nearly Gaussian \cite{Planck_2020_NG}. Nonetheless, departures from Gaussian statistics, termed primordial non-Gaussianities (PNGs), are inevitable \cite{Cabass:2016cgp} and can be substantial in a variety of scenarios, including models with strong self-interactions \cite{Babich_2004}, excited initial states \cite{Chen_2008}, or multiple light or massive fields, as in ``cosmological collider''-like models \cite{Lyth_2003,Chen:2009zp,Bodas:2020yho,McCulloch:2024hiz}. A detection of PNGs (or lack thereof) will allow us to probe these possibilities, carving out the space of microphysics present in the early Universe.

The signatures of early-Universe physics are encoded in statistical correlations of the primordial potential $\Phi$, which is related to the matter field $\delta_m$ by a transfer function that scales as $\mathcal{M}(k) \propto k^2$ on large scales, i.e.\ $\delta_m(\mathbf{k},z) = \mathcal{M}(k,z)\Phi(\mathbf{k})$. The bispectrum, $B=\langle \Phi(\mathbf{k}_1)\Phi(\mathbf{k}_2)\Phi(\mathbf{k}_3)\rangle$, which probes the skewness as a function of scale, is often the leading measure of PNGs. A class of phenomenologically interesting PNGs has bispectra with a squeezed limit ($k \gg k_\ell\to 0 $) \cite{Schmidt:2012ky,Shiraishi:2013vja,Assassi:2015jqa}
\begin{align}\label{eq:squeezedfnl}
B(\mathbf{k}_\ell,\mathbf{k},-\mathbf{k})=2 \sum_{L\text{ even}} f_\text{NL}^{s=L}\mathcal{L}_L(\hat{\mathbf{k}}_\ell\cdot\hat{\mathbf{k}}) P(k_\ell)P(k)\,,
\end{align}
where $\mathcal{L}_L$ is the Legendre polynomial of order $L$, $f_\text{NL}^{s=L}$ the corresponding PNG amplitude and $P(k)$ is the power spectrum. The local model \cite{Komatsu:2001rj} featuring $f_\text{NL}^{s=0}\equiv f^{\rm loc.}_{\rm NL}$ is a prime target for many PNG searches in large-scale structure surveys \cite{2009arXiv0912.0201L,SO_2019,Schlegel_2022,Besuner_2025,Bock_2026}, as it alters clustering of biased tracers in redshift space to 
\begin{align}\label{eq:fnl-local}
    \delta_t(\mathbf{k}) = \left(b_1+ f\mu^2 + \frac{b_{\Phi}f^{\rm loc.}_{\rm NL}}{\mathcal{M}(k)}\right)\delta_m(\mathbf{k})\,,
\end{align}
yielding a distinctive $\mathcal{M}(k)^{-1}\propto1/k^2$ enhancement (or suppression) in the clustering at large scales \cite{Dalal:2007cu,Matarrese:2008nc,Slosar:2008hx,Schmidt:2010gw,Assassi:2015fma,MoradinezhadDizgah:2017szk}.  Here, $b_1$ is the  linear bias, $f$ the logarithmic growth rate, and $\mu =\hat{\mathbf{k}}\cdot\hat{\mathbf{n}}$ is the projection of the Fourier mode along the line-of-sight (LOS) $\hat{\mathbf{n}}$. This enhancement is due to the non-Gaussian modulation of local peak heights in the presence of a long-wavelength mode as prescribed by $f^{\rm loc.}_{\rm NL}$ in \cref{eq:squeezedfnl}, giving rise to a scale-dependent bias with a response strength $b_\Phi$. This effect is known to be absent in the higher multipoles $f_\text{NL}^{s\geq 2}$, which only make collapse thresholds anisotropic \cite{Schmidt:2015xka}. The intuition is that the probability to form halos in a region of size $R$ only depends on local variances smoothed at that scale. Anisotropy solely shifts power from one direction to another, and thus does not modify abundances at linear order \cite{Schmidt:2015xka,Assassi:2015fma}. 

Hereafter, we focus our attention on $f_\text{NL}^{s=2}$, which naturally arises from higher-spin fields \cite{Arkani-Hamed:2015bza,Lee:2016vti,Bordin:2018pca}. PNGs of this type can also shed light on the vacuum state \cite{Agullo:2012cs,Mirbabayi:2022cbt}, and alternative symmetry-breaking patterns \cite{Endlich:2012pz} during inflation. For $f_\text{NL}^{s=2}$, it has been established that the modulation in small-scale matter fluctuations is quadrupolar, i.e.\ it couples local tidal fields to the long-wavelength tidal field they sit atop \cite{Schmidt:2015xka,Assassi:2015jqa,Assassi:2015fma}. This prompted explorations into finding scale-dependent bias in galaxy shape correlations \cite{Schmidt:2015xka,Akitsu-2021,Kurita-2023}, which directly probes the tidal field. Extending this philosophy, we argue that scale-dependent bias can also be made manifest if tracers were selected, by design or by virtue of their astrophysical origin, on properties linked to the tidal field, vastly expanding the scope of $f_\text{NL}^{s=2}$ searches.

In \cref{eq:fnl-local}, the coefficient in the angular dependence is fixed by the Kaiser relation. In the presence of selection effects however, some assumptions that underpin that simplicity are violated. For an ensemble of tracers selected via properties that correlate with the tidal field, their clustering would be modulated accordingly by the projection of $\mathcal{D}_{ij}\Phi\equiv (\partial_i\partial_j\nabla^{-2} - \delta_{ij}/3)\Phi$ along $\hat{\mathbf{n}}$ \cite{Assassi:2015fma}. On large cosmological scales, this can be parameterized as \cite{Mirbabayi:2014zca,Assassi:2015fma,Desjacques:2018pfv}
\begin{align}\label{eq:linearBiasWScaleDep}
\delta_t(\mathbf{k})&= (b_1+ b_\eta f\mu^2)\delta_m(\mathbf{k}) + f_\text{NL}^{s=2} b_{\Phi,\eta}\hat{n}^i\hat{n}^j\mathcal{D}_{ij} \Phi(\mathbf{k}) \nonumber \\
& = \left(b_1+b_\eta f \mu^2+\left[\mu^2-\frac{1}{3}\right]\frac{f^{s=2}_\text{NL}b_{\Phi,\eta}}{\mathcal{M}(k)} \right) \delta_m(\mathbf{k})\,.
\end{align}
In contrast to \cref{eq:fnl-local}, there are two new parameters $b_\eta\neq 1$ and $b_{\Phi,\eta}$ that accompany a $\mu$-dependence in this bias expansion, which quantify the response of the tracer to line-of-sight-dependent effects \cite{Desjacques:2018pfv}. Tracers without selection effects have $b_\eta=1$ and have zero primordial response $b_{\Phi,\eta}=0$; whilst those with them can deviate significantly from $b_\eta=1$. The dependence on the local tidal field in the selection adds a response to anisotropic collapse and, for universes with anisotropic PNGs, the squeezed coupling in Eq. \ref{eq:squeezedfnl} induces a modulation of the observed field by the large-scale tidal field. Thus, these samples can show strong signatures of $f^{s=2}_\text{NL}$-induced scale-dependent bias.

 We explore two well-studied cases where the Kaiser relation is broken with $b_\eta\neq 1$. One scenario is for observables like the observed flux of emission or absorption lines formed through spectroscopic transitions in the intergalactic medium. Here, the line-widths can be correlated with large-scale fields and complicated by radiative transfer effects \cite{Zheng:2010jf,Mao:2011xp,Irsic:2018hhg,Byrohl:2019ypo} such that they require the inclusion of $b_\eta$ as a free bias parameter. For example, this is an industry standard for analyses of the Lyman-$\alpha$ (\lya) forest \citep{DESILyaDR1BAO2025,DESILyaDR2AP2026}. Another case is when the visibility of objects, and hence their selection, depends upon their viewing angle. The observed distribution is thus conditioned on galaxy shapes, which are correlated with cosmological tidal fields. This effect was proposed in \cite{Hirata:2009qz} and detected in data in \cite{Martens:2018uqj,Obuljen:2020ypy,Lamman:2023tlr}. Whilst this effect is typically small and the viewing angle is fixed by the observer's LOS, precise spectroscopic measurements allow us to infer other galaxy properties that are correlated with the local tidal field. These measurements can in turn be used to engineer samples that maximize these selection effects, thereby allowing constraints on $f^{s=2}_\text{NL}$. 

To demonstrate, we consider the application of the ideas above to two samples and seek a $1/k^2$ scale-dependence in the power spectrum at large scales. To simplify the $\mu$-dependence, we make use of power spectrum multipoles defined as
\begin{align}
P_L(k)=\frac{2L+1}{4\pi}\int d\Omega_{\mathbf{k}}~\mathcal{L}_L(\mu)\langle \delta(\mathbf{k})\delta^*(\mathbf{k})\rangle\,.
\end{align}
Specifically, we expect the monopole signal to be
\begin{align}\label{eq:pk_auto_L=0}
\begin{split}
P^{tt}_{L=0}&=\Bigg[b_1^2 +\frac{2}{3}b_1 b_\eta f +\frac{1}{5}(b_\eta f)^2 \\ & + \frac{4}{45}\frac{f^{s=2}_\text{NL}b_{\Phi,\eta}}{\mathcal{M}(k,z)}\left(\frac{f^{s=2}_\text{NL}b_{\Phi,\eta}}{\mathcal{M}(k,z)} + 2 b_\eta f\right) \Bigg]P_{mm}(k)\,,
\end{split}
\end{align}
the quadrupole to be
\begin{align}\label{eq:pk_auto_L=2}
\begin{split}
P^{tt}_{L=2}&=\frac{4}{63}\left(b_\eta f + \frac{f^{s=2}_\text{NL}b_{\Phi,\eta}}{\mathcal{M}(k,z)} \right) \\ 
& \times \left(21 b_1 + 9 b_\eta f + \frac{2f^{s=2}_\text{NL}b_{\Phi,\eta}}{\mathcal{M}(k,z)} \right) P_{mm}(k)\,,
\end{split}
\end{align}
and the hexadecapole to be
\begin{align}\label{eq:pk_auto_L=4}
P^{tt}_{L=4}=\left[\frac{8}{35}\left(b_\eta f +\frac{f^{s=2}_\text{NL}b_{\Phi,\eta}}{\mathcal{M}(k,z)}\right)^2\right] P_{mm}(k)
\end{align}
for the tracer autospectrum, and 
\begin{align}
P^{mt}_{L=0}=\left[b_1+\frac{1}{3}b_\eta f\right] P_{mm}(k)
\end{align}
and
\begin{align}
P^{mt}_{L=2}=\left[\frac{2}{3}\left(b_\eta f+\frac{f^{s=2}_\text{NL}b_{\Phi,\eta}}{\mathcal{M}(k,z)}\right)\right] P_{mm}(k)
\end{align}
for the matter-tracer cross spectrum.

\begin{figure}
\includegraphics[width=.5\textwidth]{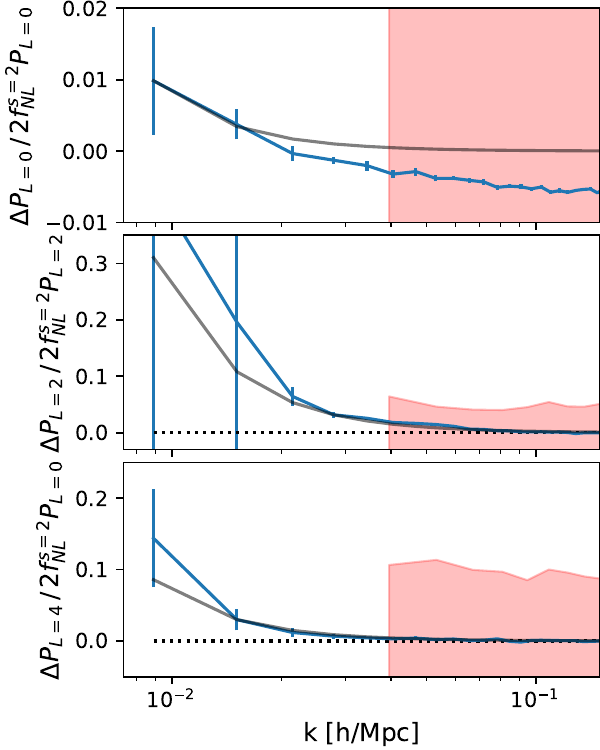}
\caption{The mock Lyman-$\alpha$ power spectrum measurements at $z=2$. The panels show the response of the power spectrum to $f_\text{NL}^{s=2}$ divided by the $f_\text{NL}^{s=2}=0$ power spectrum for the monopole (top), quadrupole (middle) and hexadecapole (bottom). The response is computed from the difference of the statistic computed between the $f_\mathrm{NL}^{s=2}=300$ and $f_\mathrm{NL}^{s=2}=-300$ simulations and forms an approximate numerical derivative. This normalization removes the scale-dependence from the primordial power spectrum and isolates any $1/k^2$ enhancement.
On large scales, the $1/k^2$ feature, described in this paper and with the expected scaling in grey, is clearly seen. For contextualization we show the fractional error bars obtained from the eBOSS analysis in \citep{deBelsunce-2024}. The hexadecapole normalization is chosen to avoid Monte Carlo noise and is further discussed in Sup. Mat. Sec I.
\label{fig:pk_lya_all}}
\end{figure}

\textit{Results I: Lyman-$\alpha$ Forest.} --- 
The {\lya} forest  describes the features imprinted into the spectra of high redshift objects through scattering off neutral hydrogen. It is well known that this observable has $b_{\eta}\neq1$ \citep{Croft-1998,McDonald-2000}. This arises due to the non-linear mapping between the optical depth and the observed Ly$\alpha$ flux \citep{Seljak-2012} and also due to correlations between large-scale velocity flows and small-scale velocities that set the absorption profile width \citep{Arinyo-2015,Irsic:2018hhg}. Thus, the Ly$\alpha$ forest is a natural case to search for $f_{\rm NL}^{s=2}$.

To study this, we generate a set of simulated Ly$\alpha$ observations in a universe with and without $f_{\rm NL}^{s=2}$. These N-body simulations use standard methods: the initial conditions were generated according to Ref. \cite{Akitsu-2021}, simulations with $1280^3$ dark matter particles in a volume of $1$ Gpc$^3$ were run with \textsc{pkdgrav3} \cite{Potter-2017} and the observations were created with the fluctuating Gunn-Peterson approximation (FGPA) \cite{Croft-1998} at $z=2.0$. Details of the implementation are described in Sup. Mat. Sec. I.

Measurements of the power spectrum multipoles of this field are shown in \cref{fig:pk_lya_all}. The simulation clearly demonstrates our new signal. To contrast with the usual local-type non-Gaussianity,  although $f_{\rm NL}^{\rm loc.}$ generates a scale-dependence in the monopole and quadrupole as well; the relative strengths are different and no signal is expected in the hexadecapole. The exponential optical-depth-to-flux mapping introduces strong non-linearities. Consequently, $b_{\Phi,\eta}$ for {\lya} is naturally of the order $\mathcal{O}(1)$, explaining the prominence of the $1/k^2$ signal in the multipoles of the mock {\lya} measurements. The amplitude suppression on all scales in the monopole arises due to non-linearities in the FGPA and in the mapping mentioned above. A similar effect is also present in $f^{\rm loc.}_\mathrm{NL}$ simulations. For comparison, we include error bars from analysis of the eBOSS dataset \citep{deBelsunce-2024} in \cref{fig:pk_lya_all}. Future {\lya} forest measurements will be very constraining, though observational systematic effects due to quasar continuum treatment of large-scale measurements \citep{deBelsunce-2024} limit the $k_{\rm min}$ of the current analysis.

\begin{figure}
\includegraphics[width=.5\textwidth]{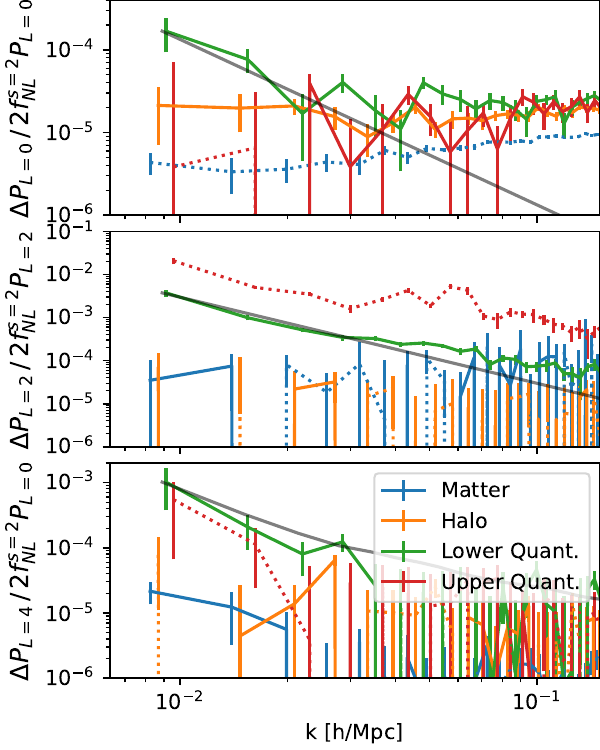}
\caption{The response to $f_\mathrm{NL}^{s=2}$, as in  \cref{fig:pk_lya_all}, for the auto power spectrum multipoles for the matter field (blue), full halo sample (orange), the upper quantile sample (green) and the lower quantile sample (red). The top panel shows the monopole, the middle panel the quadrupole and the bottom panel shows the hexadecapole. The grey line denotes the predicted $1/k^2$ scaling and the dotted lines show negative signals.
\label{fig:pk_sel_auto_all}}
\end{figure}
\begin{figure}
\includegraphics[width=.5\textwidth]{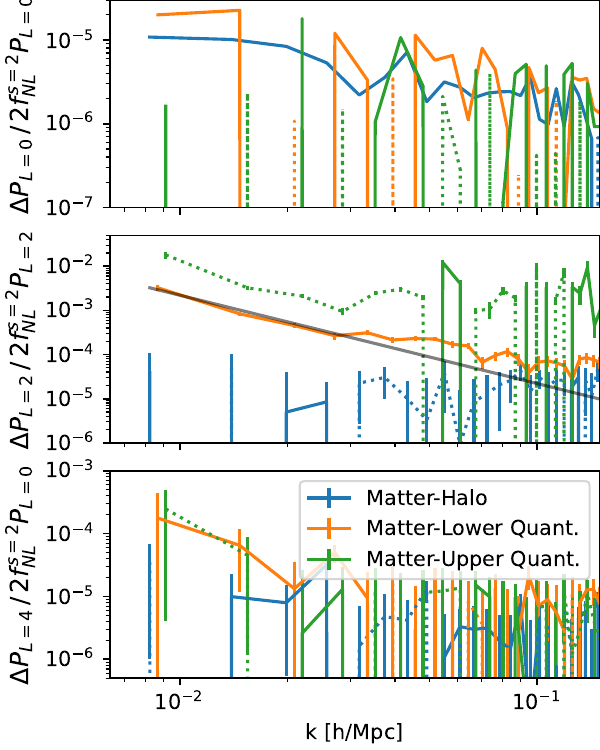}
\caption{The response to $f_\mathrm{NL}^{s=2}$, as in  \cref{fig:pk_lya_all}, for the halo-matter power spectrum multipoles for the three samples at $z=2.0$: the full halo sample (with mass $M>4.3\times10^{12}M_\odot/h$), and the two samples selected on their radial size. The panels show the approximate derivative with respect to $f_\text{NL}^{s=2}$ of the logarithm on the power spectrum monopole (top), quadrupole (middle) and hexadecapole (bottom). Whilst the full sample, which has $b_\eta=1$, does not respond strongly, the selected tracers exhibit a strong, $1/k^2$ response in the quadrupole. The grey line denotes the predicted $1/k^2$ scaling and the dotted lines show negative signals. \label{fig:pk_sel_cross_all}}
\end{figure}

\textit{Results II: Galaxy Clustering.} --- 
Selection effects in 3D galaxy clustering can lead to galaxy samples with $b_\eta\neq1$. We demonstrate this with our simulations using dark matter halos, biased tracers formed of gravitationally collapsed objects, at $z=2.0$. We form two halo samples that correspond to the upper 50\% and lower 50\% in their LOS size. \cref{fig:pk_sel_auto_all} and \cref{fig:pk_sel_cross_all} clearly show the $1/k^2$ enhancement from $f_{\rm NL}^{s=2}$ in the halo power spectrum and the cross spectrum with the dark matter density. Importantly, the signal is absent in the matter field and the full halo sample, for which the Kaiser relation holds and forces $b_\eta=1$. Local PNG does not generate a scale-dependent signal in the hexadecapole of the auto spectrum nor in the quadrupole of the cross spectrum, further demonstrating a clear signature of our model. An interesting point to note is that the two samples have different response strengths to $f^{s=2}_\mathrm{NL}$, i.e.\ different $b_{\Phi,\eta}$.

Having demonstrated this concept on simulations, we proceed to analyze the Baryon Oscillation Spectroscopic Survey (BOSS) Data Release (DR) 12 dataset obtained from a spectrograph at the Apache Point Observatory \cite{Dawson-2013,Alam-2017}. The BOSS dataset contains catalogs of 3D galaxy positions in redshift space, and a set of galaxy properties obtained from detailed analysis of their spectra. We use two samples selected by their stellar mass and velocity dispersion that Ref. \cite{Obuljen:2020ypy} designed for the LOWZ North and CMASS North observations. These samples select objects that are roughly in the upper and lower quartile of the observed radial velocity dispersion (hereafter upper and lower quant. samples). As velocity dispersions are primarily generated through gravitational collapse, the velocity dispersion tensor $\sigma_{ij} \equiv \langle v_i v_j\rangle - \langle v_i \rangle\langle v_j\rangle$ largely traces the anisotropies set by the local tidal field \cite{Ragone-Figueroa_2010,Aviles:2015osc,Buehlmann:2018qmm,Obuljen-2019}. The radial velocity dispersion, as measured through emission line widths, thereby provides a proxy field that enables selection of the projected tidal field. The specific construction uses the stellar mass to ensure that the linear bias of the two samples is similar and we refer the reader to Ref. \cite{Obuljen:2020ypy} for further details.

\begin{figure}
\includegraphics[width=.5\textwidth]{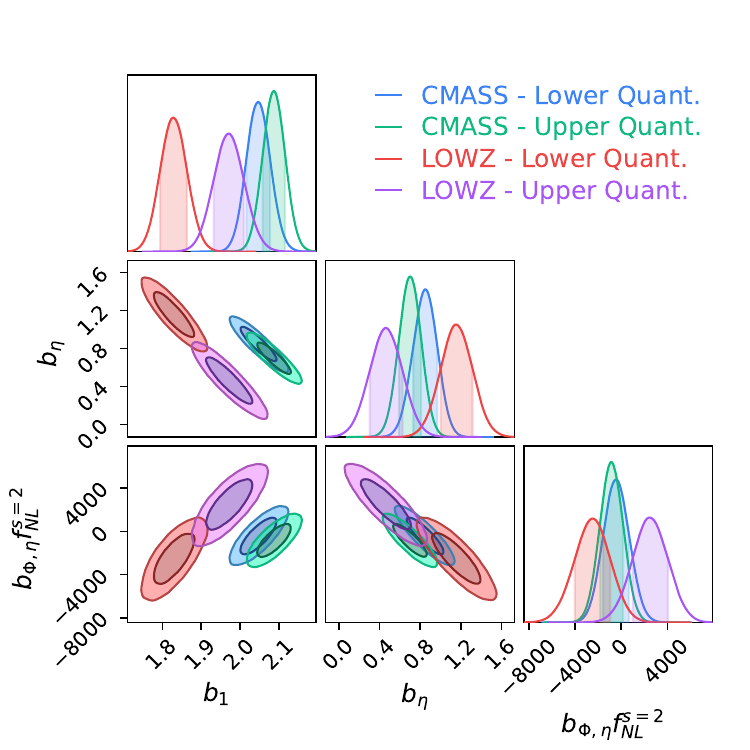}
\caption{Constraints on the BOSS samples using our linear model. First note that the $b_\eta$ of the samples is different, most evidently for LOWZ, whilst the $b_1$ values are similar. Second, our measurements of $b_{\Phi,\eta}f_\mathrm{NL}^{s=2}$ are consistent with zero, thus we have no detection of primordial non-Gaussianity.
\label{fig:constrains_multisample}}
\end{figure}

We analyze the 3D power spectrum multipoles of this dataset with the \textsc{polybin3d} code\footnote{https://github.com/oliverphilcox/PolyBin3D}\cite{Philcox-2021,Philcox-2025}, which has been used to analyze BOSS and DESI datasets \cite{Ivanov-2023,Chudaykin-2025}.  Details of these measurements and our analysis choices are described in Sup. Mat. Sec. II. For each sample, we assume a Gaussian likelihood for the data and fit the linear model given in \cref{eq:linearBiasWScaleDep}. The results of these fits are shown for all our samples in \cref{fig:constrains_multisample} and are good fits to the data with all Probability-to-Exceed values between $0.11$ and $0.77$. We find no statistically significant evidence for $f_{\rm NL}^{s=2}$. Through galaxy shape correlations, Ref. \cite{Kurita-2023} found $-62.4<(b_{\Phi,\eta} f_\mathrm{NL}^{s=2})\leq 15.4$ at 68\% confidence. Whilst this is a significantly tighter constraint, the expected value of $b_{\Phi,\eta}$ for galaxy shapes is of order $\mathcal{O}(10^{-2})$ \citep{Akitsu-2021,Kurita-2023}. In our simulated halo samples, the selection effect sample bias (checked at $z=0.5$) is of order $\mathcal{O}(1)$, which leads to competitive constraints on $f_\mathrm{NL}^{s=2}$. We discuss this further in the conclusions.

\textit{Conclusions.} --- 
This paper presents a new method to constrain early-Universe physics via a new scale-dependent feature. This feature arises when effects break the symmetry requirements that typically fixes the angular dependence in the clustering of large-scale structure tracers. This feature allows models of inflation that predict PNGs with an angle-dependent coupling between long and short modes (known as anisotropic PNG) to be constrained with standard power spectrum tools, rather than complex and computationally expensive bispectrum measurements. We demonstrate this effect on simulated observables, and constrain the anisotropic PNG amplitude $f_\mathrm{NL}^{s=2}$ with BOSS galaxy measurements. We find no evidence for this type of primordial physics.

The data analysis presented here demonstrates the sensitivity of our method; however, there are numerous interesting extensions that would improve the constraining power. Firstly, joint analysis of two samples of the same dark matter field allows the bias parameters to be measured without  ``noise''  by sampling the variance of the matter field. Sample variance cancellation has been demonstrated to be a powerful tool \cite{Seljak-2009,Schmittfull-2018}. We did not apply it here as our samples have non-trivial correlations that would need to be accounted for. For example, the alignment of shapes and velocity dispersion follows the tidal field and so objects in the two samples will typically be separated by the tidal field coherence length. Second, we did not combine any of the samples due to each of them having different bias coefficients that are unknown. Methods have been developed to model and measure the equivalent coefficients for local PNG \cite{Barreira-2020,Sullivan-2025,Dalal-2026}. It would be highly beneficial to transfer these ideas to anisotropic PNG. Finally, in this work we chose to limit the analysis to large scales where a linear model is accurate. Smaller-scale measurements would help break degeneracies and improve precision, as we demonstrate in  Sup. Mat. Sec. II. However, to perform such an analysis robustly, a better theoretical model is necessary. EFT models have been highly successful in addressing this challenge; however existing models cannot be directly used in our case as additional terms will need to be added to account for anisotropic PNG (analogous to those in Refs. \cite{Assassi:2015jqa,Assassi:2015fma,Cabass-2022,DAmico-2025}) in a framework that fully encapsulates selection-dependent effects, such as those explored in \cite{Desjacques:2018pfv,Ivanov-2024,deBelsunce-2024,Ivanov-2025}.

Taken together, this work opens a new avenue of research on constraining anisotropic PNG.

\acknowledgments
\textit{Acknowledgments:} \small{WRC is supported by the Science and Technology Facilities Council (STFC) through an Ernest Rutherford Fellowship (UKRI2424: ``Astrophysics and Cosmology from CMB Scattering'') and through the Royal Society grant ICA\textbackslash R2\textbackslash 252140.  YSY is supported by the Isaac Newton Studentship of the Cambridge Trust. The authors are grateful to Oliver Philcox, Elisa Chisari and Azadeh Moradinezhad Dizgah for insightful discussions. } 
\section{Supplementary Material}
 
\section{I. Simulations \label{sec:sims}}
To validate this effect, we generate a set of cosmological simulations with anisotropic primordial non-Gaussianity. We largely follow the methods used in Ref. \cite{Akitsu-2021} and so provide an overview here and refer the reader to that reference for details. 

\begin{figure}
\includegraphics[width=.5\textwidth]{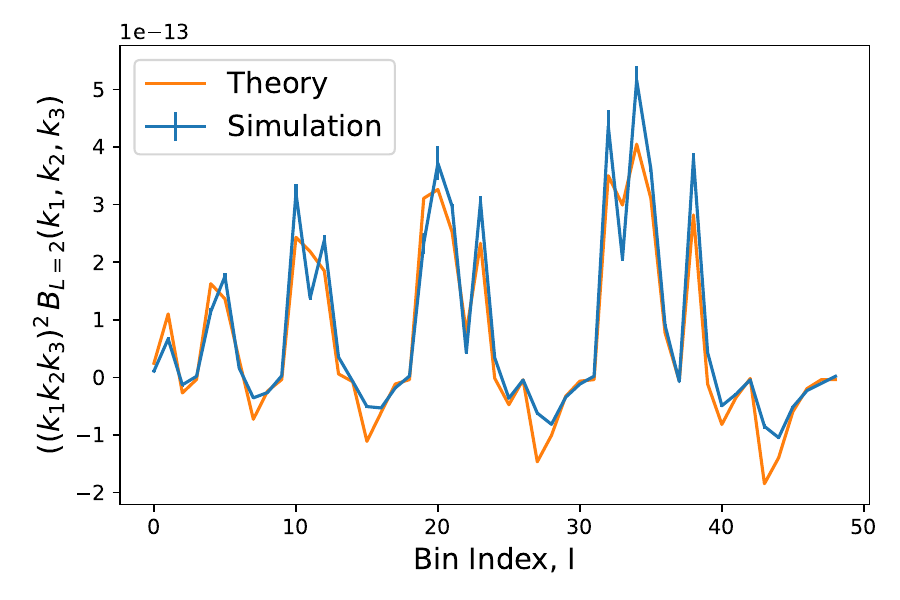}
\caption{A comparison of the bispectrum quadrupole  of the initial conditions to the theoretical expectation. This provides a validation of the implementation of our initial conditions. \label{fig:ic_bispec}}
\end{figure}

First, we generate a primordial Gaussian potential field $\Phi^G(\mathbf{x})$. We generate primordial non-Gaussianity with the following operation
\begin{align}
\Phi^{NG}(\mathbf{x})=\Phi^G(\mathbf{x})+\frac{2}{3}f^{s=2}_\text{NL}\sum_{ij}\left[\Phi_{ij}(\mathbf{x})^2-\langle \Phi_{ij}^2\rangle \right]
\end{align}
where
$\Phi_{ij}(\mathbf{x})=\frac{3}{2}\left[\frac{\partial_i\partial_j}{\nabla^2}-\frac{1}{3}\delta_{ij}\right]\Phi^G(\mathbf{x}).$ This generates a primordial bispectrum of the form of \cref{eq:squeezedfnl}. We use a modified version of 2LPTIC as developed in \cite{Scoccimarro-2012} and extended in \cite{Coulton-2023}.
We compute the binned bispectrum quadrupole of the initial conditions, defined as
\begin{align}
B_{L=2}(k_1,k_2,k_3) = &\frac{1}{2}\left[3 (\hat{\mathbf{k}}_1\cdot\hat{\mathbf{k}}_2)^2-1\right] \Phi(\mathbf{k}_1) \Phi(\mathbf{k}_2) \Phi(\mathbf{k}_3)\nonumber \\& + \mathrm{ permutations},
\end{align}
averaged over modes within 6 equally spaced k-bins between $k_\mathrm{min}=0.01$h/Mpc and $k_\mathrm{ max}=0.15$h/Mpc.
In \cref{fig:ic_bispec} we compare these bispectrum measurements with the theoretical prediction. The reasonable agreement provides a validation of our implementation.

This primordial field is evolved to $z=127$ with 2LPT and then the resulting field is evolved to $z=2$ using the \textsc{pkdgrav3} N-body code \cite{Potter-2017}. We simulate a 1 Gpc$^3$ box with 1280$^3$ dark matter particles at the Quijote fiducial cosmology \cite{Villaescusa-Navarro-2020}. We generate 20 realizations of $\Phi^G$ and for each one we run three N-body simulations with $f^{s=2}_\text{NL}=-300,0,$ and $300$. These matched-phase runs allow us to minimize the impact of cosmic variance. Finally, we run the \textsc{rockstar} halo finder on the snapshot to identify dark matter halos \cite{Behroozi-2013}.  We make two types of simplified mock observables: a sample of tracers with a strong anisotropic selection bias and a set of mock Lyman-$\alpha$ forest lines.

 For the selection bias tracers, we use the halo catalogs and split them into three samples: the ``full sample'' of all halos with $>100$ dark matter particles, the ``lower quantile sample''  whose shape in the observed direction (taken to be the x-axis) is below the median size and the ``upper quantile sample'' whose shape in the observed direction is above the median. We use \textsc{rockstar}'s weighted shape tensor to calculate the shape. This approach mirrors that described in Ref. \cite{Obuljen-2019}.

To make mock Lyman-$\alpha$ forest lines we use the fluctuating Gunn Peterson approximation (FGPA) \cite{Croft-1998} and we closely follow the implementation in \cite{Hadzhiyska-2023}. The FGPA assumes that the baryons trace the dark matter, that adiabatic expansion of the gas leads to a power law equation of state for the gas $T = T_0(1+\delta)^{\gamma-1}$ and that the balance of photo-ionization and collisional recombination produces an optical depth $\tau$ as
\begin{align}
\tau(\mathbf{x})=\tau_0 (1+\delta(\mathbf{x}))^2T(\mathbf{x})^{-0.7}.
\end{align}
We assume $\gamma=1.5$. We then apply redshift space distortions accounting for the thermal broadening of the line with a Gaussian of width $b=\sqrt{2k_B T(\mathbf{x})/m_p}$, where $k_B$ is the Boltzmann constant and $m_p$ is the proton mass. This is finally converted to the observable transmitted flux fraction as $F=\exp(-\tau)$. The mean optical depth is set by matching the mean flux from Ref. \cite{Faucher-Giguere-2008}. Note we compute the optical depth on a grid, compute per-particle optical depth and temperature and then apply the RSD to the particles using the dark matter particle velocities as well as a thermal broadening kernel.

Once we have simulated the observables, we then compute the power spectrum of these fields and the cross correlation with the matter field. We use the \textsc{pylians3} code\footnote{https://github.com/franciscovillaescusa/Pylians} to compute these on $256^3$ grids with a Cloud-in-Cell (CiC) kernel to assign particles to the grid.

The measurements reported as responses are computed using the difference of the simulations at $f^{s=2}_\mathrm{NL}=300$ and $f^{s=2}_\mathrm{NL}=-300$. This is divided by the statistic computed from simulations with $f_{\mathrm{NL}}^{s=2}=0$ to remove the scale-dependence from the primordial power spectrum and this isolates any large-scale scale-dependent bias. Note that for the hexadecapole measurement we divide by the monopole as the hexadecapole is very noisy on these scales and the noise fluctuations around zero cause a diverging division. \cref{eq:pk_auto_L=0,eq:pk_auto_L=4} show that this still isolates the scale-dependent bias terms.

\section{II. BOSS Measurement}\label{sec:BOSS_measuremet}
In this section we provide details of the sample selection, the power spectrum estimation pipeline, a discussion of the measurements and scale cuts.

\subsection{Sample Selection}
We use the BOSS DR12 catalogs described in \cite{Reid-2016} and from the extended Portsmouth galaxy property catalogs \footnote{https://www.sdss4.org/dr12/spectro/galaxy\_portsmouth/} we use the velocity dispersions and stellar masses (specifically the passive model with the Kroupa initial mass function), derived using the methods from Ref. \cite{Thomas-2013} and Ref. \cite{Maraston-2013} respectively. Our analysis focuses just on the North part of the BOSS observations.  The generation of the catalogs follows Ref. \cite{Obuljen:2020ypy}. We outline the procedure here and refer the reader to  Ref. \cite{Obuljen:2020ypy} for a detailed discussion of the choices.  

The first step is to divide each of the LOWZ and CMASS observations into 30 evenly spaced redshift bins. Then we compute the quantile value of each galaxy's velocity dispersion and stellar mass. Two boundaries in velocity dispersion quantile ($\sigma_\star[\%]$) vs stellar mass quantile  ($ M_\star[\%]$) are defined using the following relation:
\begin{align}
    \sigma_\star[\%] = A \times M_\star[\%]+B.
\end{align}
The parameters A and B are obtained from Table A in Ref. \cite{Obuljen:2020ypy}. We do not use the aperture correction. The upper boundary defines a sample where galaxies with $\sigma_\star[\%]> A_\mathrm{upper} \times M_\star[\%]+B_\mathrm{upper}$ and the second boundary defines a sample where galaxies with $\sigma_\star[\%]< A_\mathrm{lower} \times M_\star[\%]+B_\mathrm{lower}$. Hereafter we refer to these as the upper and lower quant. samples respectively and each contains approximately 25\% of the total sample. This procedure was designed to have samples that approximately preserve the redshift and spatial selection of the full sample and so that the two samples have approximately the same linear bias but different $b_\eta$.  We generate random catalogs for these by subsampling the BOSS random catalogs. Following Ref. \cite{Obuljen:2020ypy} we use shuffled redshifts of the true galaxies for the redshifts of our subsamples.

\subsection{Measurement Pipeline}
This pipeline follows the steps used in Ref. \cite{Chudaykin-2025}. This ``unwindowed'' approach has been thoroughly tested and allows for an accurate computation of the mask coupling terms, the integral constraints, the covariance matrices and optimal weighting schemes (the later is not used here). We outline the key steps and refer the reader to Ref. \citep{Philcox-2021} for more details.

First we convert our catalogs to 3D spatial points assuming a \textit{Planck}-like cosmology; we do not account for the Alcock-Paczynski effect \cite{Alcock-1979}. We compute the completeness weights, defined as $w_c=w_{\rm sys}(w_{\rm rf}+w_{\rm fc}-1)$ where the terms are the systematic, redshift failure and fiber collisions weights, and the updated Feldman, Kaiser, and Peacock (FKP) weights \cite{Feldman-1994}, assuming $P_0=10^4$ $({\rm Mpc}/h)^3$. Next we generate grids of the observations and randoms and we compute the background density, $n(\mathbf{x})$ as 
$\alpha$ times the random field, where  $\alpha=\sum_\mathrm{data} w_i/\sum_\mathrm{randoms} w_i$. Our input data vector, $d(\mathbf{x})$, is computed as the difference between the observed galaxy field and the background density, with both fields weighted by the combined FKP and systematic weights. The final field we compute is the doubly weighted mask field defined as 
\begin{align}
    n_2(\mathbf{x})=(\alpha_2+\alpha^2)\sum_\mathrm{randoms} w^2_{i}\delta^{(3)}(x-x_i),
\end{align}
where $\alpha_2=\sum_\mathrm{data} w_i^2/\sum_\mathrm{randoms} w_i^2$.

Next, we compute the unwindowed power spectrum, $\hat{P}_{L}(k)$. This is done in three steps. First, we computed the data quadratic estimate defined as
\begin{align}
      &\hat{P}^{qe}_{L}(k_a) =\frac{1}{2}\int\frac{\mathrm{d}^3\mathbf{k}}{(2\pi)^3}\theta_{k_a}(k)\nonumber \\ & \times \int\mathrm{d}^3\mathbf{x} e^{-i\mathbf{k}\cdot{\mathbf{x}}}\mathcal{L}_L(\hat{\mathbf{k}}\cdot\hat{\mathbf{x}})M^{-1}d(\mathbf{x}) \int\mathrm{d}^3\mathbf{y} e^{i\mathbf{k}\cdot{\mathbf{y}}}M^{-1}d(\mathbf{y})
\end{align}
where $\theta_{k_a}(k)$ is a top hat function that is one only where the $k$ value falls within the bin, and $M$ is the pixel window function arising from the grid assignment operation (in our case a CiC window) and this is used to remove all the CiC gridding windows. Next we compute the shot noise bias and subtract this from the quadratic estimate. The shot noise bias is defined as
\begin{align}
    \hat{P}^{sn}_{L}(k) = \frac{1}{2}\int\mathrm{d}^3\mathbf{x} \int\frac{\mathrm{d}^3\mathbf{k}}{(2\pi)^3}\theta_{k_a}(k)\mathcal{L}_L(\hat{\mathbf{k}}\cdot\hat{\mathbf{x}})n_2(\mathbf{x}).
\end{align}
Finally we compute the Fisher matrix as
\begin{align}\label{eq:fisherPk}
  &  \mathcal{F}_{(L,k_a),(L',k_b)}=\nonumber \\&\frac{1}{2}\int\mathrm{d}\mathbf{x}\mathrm{d}\mathbf{y}\mathrm{d}\mathbf{z}\mathrm{d}\mathbf{w}
  \int
  \frac{\mathrm{d}^3\mathbf{k}}{(2\pi)^3}\theta_{k_a}(k)\mathcal{L}_L(\hat{\mathbf{k}}\cdot\hat{\mathbf{x}})e^{-i\mathbf{k}\cdot({\mathbf{x}-\mathbf{y})}}P(y,z)  \nonumber \\&
  \times\int
  \frac{\mathrm{d}^3\mathbf{k'}}{(2\pi)^3}\theta_{k_b}(k')\mathcal{L}_{L'}(\hat{\mathbf{k'}}\cdot\hat{z})e^{-i\mathbf{k'}\cdot(\mathbf{z}-\mathbf{w})} P(x,w) 
\end{align}
where the pointing matrix accounts for the radial integral constraint and is given by
\begin{align}
P(\mathbf{x},\mathbf{y})=n(\mathbf{x})\left[
\delta^{(3)}(\mathbf{x}-\mathbf{y}) - \frac{n_{\rm IC}(\mathbf{y})\epsilon(x,y)}{\int \mathrm{d}\mathbf{z}\,n_{\rm IC}(\mathbf{z})\epsilon(y,z)}
\right]
\end{align}
with $n_{\rm IC}$ the background density computed without the FKP weights and $ \epsilon(x, y) = 1$ if $x$ and $y$ are in the same radial bin and zero otherwise. This is computed using Monte-Carlo estimation. The unwindowed estimate is computed by applying the inverse of the Fisher matrix to the shot noise subtracted quadratic estimate.

\begin{figure}
\includegraphics[width=.5\textwidth]{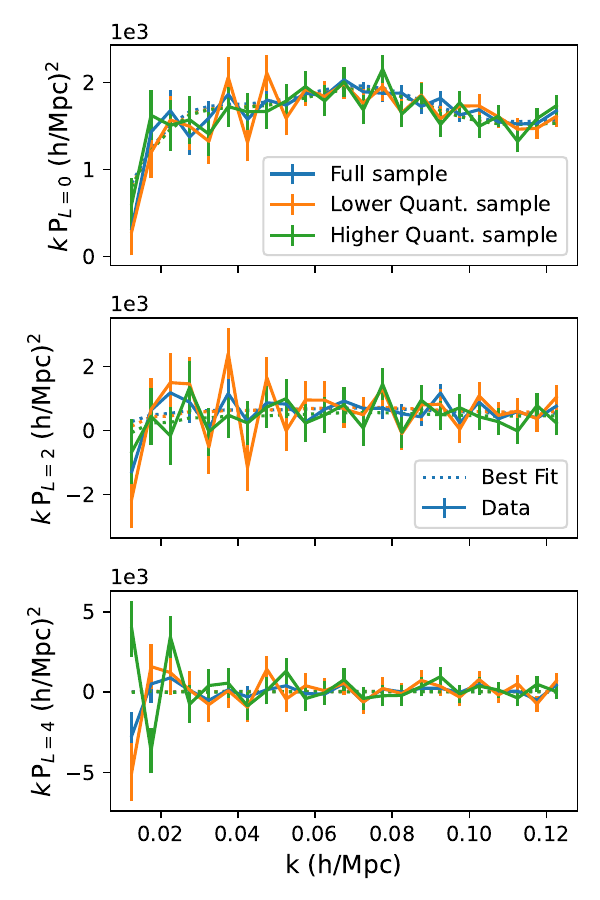}
\caption{Measurements of the power spectrum multipoles for the CMASS North Obuljen et al subsamples. We show the best fit models in dotted lines. \label{fig:CMASS_pk}}
\end{figure}

\begin{figure}
\includegraphics[width=.5\textwidth]{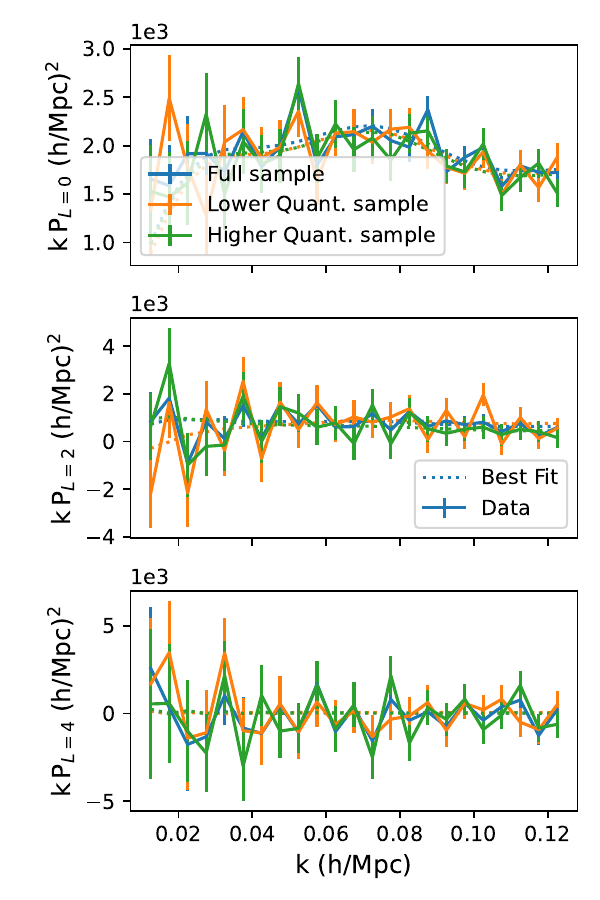}
\caption{Measurements of the power spectrum multipoles for the LOWZ North of Obuljen et al subsamples.  We show the best fit models in dotted lines.\label{fig:LOWZ_pk}}
\end{figure}

The measurements obtained from this pipeline are shown in \cref{fig:CMASS_pk} and \cref{fig:LOWZ_pk}. 

\subsection{Likelihood, model choices and scale cuts}
We use a Gaussian likelihood, $\mathcal{L}_G$, for our inferences as \begin{align}
   \ln \mathcal{L}_G = -\frac{1}{2}\left[\hat{P}_{L}(k)-\tilde{P}^T_{L}(k)\right]\Sigma^{-1} \left[\hat{P}_{L'}(k')-\tilde{P}^T_{L'}(k')\right].
\end{align}
We approximate the covariance matrix with just the Gaussian, disconnected terms as was used for the DESI analysis in Ref. \cite{Chudaykin-2025}. This has a simple form of 
\begin{align}
    \Sigma = \mathcal{F}^{-1}\tilde{F}\mathcal{F}^{-1}
\end{align} where $\tilde{F}$ is given by the Fisher formula, \cref{eq:fisherPk}, with both occurrences of the pointing matrix replaced by the data correlation function. The covariance matrix is estimated from Monte Carlo realizations of the Fisher matrix using \textsc{polybin3D}. The theory spectrum, $\tilde{P}^T(k)$ accounts for the residual mask coupling via 
\begin{align}
    \tilde{P}^T(k) = \sum_{k'} \mathcal{F}^{-1}\mathcal{G}(k')P^T(k')
\end{align}
where $\mathcal{G}$ computes the impact of masking on a fine binned theory spectrum ($P^T(k)$).\footnote{This term is needed as our observations measure binned power spectra that we then deconvolve the mask from. The binning and masking operators do not commute and hence our unwindowed estimators are not the same as binned theory values, though the differences are small.} This matrix is computed in an almost identical manner to the Fisher matrix in \cref{eq:fisherPk}, with the second binning operator ($\theta_{k_b}$) replaced by a very narrow bin.  We use broad flat priors and sample the likelihood with \textsc{emcee} \citep{Foreman-Mackey-2013}. The plots are generated with \textsc{chainconsumer} \cite{Hinton2016}.

Our theory model uses the linear model given in \cref{eq:pk_auto_L=0,eq:pk_auto_L=2,eq:pk_auto_L=4}. We assume a \textit{Planck} cosmology \cite{Planck-2018-VI} for the matter power spectrum and we do not fit for $\Lambda$CDM parameters in this work. Those parameters are sufficiently well constrained that their uncertainty is expected to have minimal impact on our constraint.

\begin{figure}
\includegraphics[width=.5\textwidth]{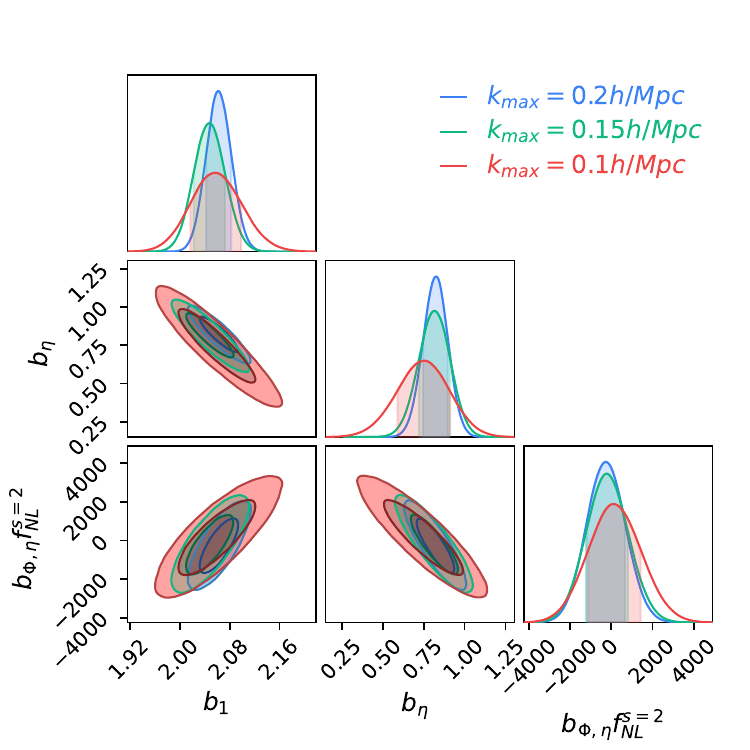}
\caption{An exploration of how varying the maximum scale, $k_\mathrm{max}$, impacts our constraining power. This demonstrates the value of small scales; however, due to our linear model, analysis with $k_\mathrm{max}>0.125$ $h/$Mpc are not trustworthy.
\label{fig:CMASS_vary_kmax}}
\end{figure}

We use a $k_\mathrm{min}=0.01$ $h/$Mpc as was adopted in \cite{Cabass-2022}, including larger scales would significantly improve our constraints, but that would require a careful study and removal of the residual large-scale systematics. We assume a $k_\mathrm{max}=0.125$ $h/$Mpc. This choice was made due to our simple, linear model. In \cref{fig:CMASS_vary_kmax}, we explore the potential value of including smaller scales. We find there are some improvements, though somewhat limited. This is expected as our signal primarily affects large scales. The improvement from increasing $k_\mathrm{max}$ arises as $b_{\eta}$ is partially degenerate with $f_{\rm NL}^{s=2}$ when only a narrow range of scales is measured.

In the past decade, effective field theory models have been effective at modeling galaxy clustering observables in the mildly non-linear regime \cite{Baumann_2012,Carrasco_2012,Carrasco_2014,Senatore_2015,Senatore_2014,Vlah_2015}. Unfortunately, the same features that produce our signal mean that existing EFT codes cannot be directly used and require extensions to accurately model our samples. 

To understand this let us consider the leading order terms. For 3D observables in real space, rotational and translational symmetry require that our observable only depends on scalar quantities. The equivalence principle and locality means that the only quantities that the density field can depend on are $\partial_i\partial_j\Phi^N$ (where $\Phi^N$ is the Newtonian potential), velocity gradients and their derivatives contracted with either $\delta_{ij}$ or with themselves. Thus, at linear order we simply have $\delta_g = b_1 \delta_m$ (as $\delta_m \propto \delta_{ij}\partial_i\partial_j\Phi^N$). The transformation to redshift space is also fixed by the equivalence principle and tracer conservation, thus we have  $\delta_g = (b_1+f\mu^2) \delta_m$. This principled approach can be extended rigorously to obtain the 1- and 2-loop predictions needed for smaller-scale analysis \cite{Angulo_2015,Baldauf_2021,Bakx_2026}. 

For our signal there is a key difference: 3D symmetry is broken to 2D rotations around the LOS. This can arise from the LOS selection, for galaxy clustering, or the LOS absorption for the Ly$\alpha$ forest. This altered symmetry group means that our observables can now depend on contractions with the line-of-sight $\hat{n}_i\hat{n}_j$, i.e. more operators are required in the EFT expansion. At the linear level this simply leads to $\delta_g = (b_1+b_\eta f\mu^2) \delta_m$ and the appearance of our signal; however, at higher order many more operators need to be included to consistently compute the 1- and 2-loop terms. This work is ongoing with Refs. \cite{Ivanov-2024,Ivanov-2025} leading the development of this framework for the Ly$\alpha$ forest. To account for $f_{\rm NL}^{s=2}$ that framework needs to be extended to include the PNG related operators, as was done for local PNG in \cite{Cabass-2022,DAmico-2025}. We leave this for future work. 

\bibliography{refs}

@article{Cabass:2016cgp,
    author = "Cabass, Giovanni and Pajer, Enrico and Schmidt, Fabian",
    title = "{How Gaussian can our Universe be?}",
    eprint = "1612.00033",
    archivePrefix = "arXiv",
    primaryClass = "hep-th",
    doi = "10.1088/1475-7516/2017/01/003",
    journal = "JCAP",
    volume = "01",
    pages = "003",
    year = "2017"
}

@ARTICLE{Chen_2008,
       author = {{Chen}, Xingang and {Easther}, Richard and {Lim}, Eugene A.},
        title = "{Generation and characterization of large non-Gaussianities in single field inflation}",
      journal = {\jcap},
         year = 2008,
        month = apr,
       volume = {2008},
       number = {4},
          eid = {010},
        pages = {010},
          doi = {10.1088/1475-7516/2008/04/010},
archivePrefix = {arXiv},
       eprint = {0801.3295},
 primaryClass = {astro-ph},
       adsurl = {https://ui.adsabs.harvard.edu/abs/2008JCAP...04..010C}
}

@ARTICLE{Lyth_2003,
       author = {{Lyth}, David H. and {Ungarelli}, Carlo and {Wands}, David},
        title = "{Primordial density perturbation in the curvaton scenario}",
      journal = {\prd},
         year = 2003,
        month = jan,
       volume = {67},
       number = {2},
          eid = {023503},
        pages = {023503},
          doi = {10.1103/PhysRevD.67.023503},
archivePrefix = {arXiv},
       eprint = {astro-ph/0208055},
 primaryClass = {astro-ph},
       adsurl = {https://ui.adsabs.harvard.edu/abs/2003PhRvD..67b3503L}
}

@ARTICLE{Babich_2004,
       author = {{Babich}, Daniel and {Creminelli}, Paolo and {Zaldarriaga}, Matias},
        title = "{The shape of non-Gaussianities}",
      journal = {\jcap},
         year = 2004,
        month = aug,
       volume = {2004},
       number = {8},
          eid = {009},
        pages = {009},
          doi = {10.1088/1475-7516/2004/08/009},
archivePrefix = {arXiv},
       eprint = {astro-ph/0405356},
 primaryClass = {astro-ph},
       adsurl = {https://ui.adsabs.harvard.edu/abs/2004JCAP...08..009B}
}

@article{Bodas:2020yho,
    author = "Bodas, Arushi and Kumar, Soubhik and Sundrum, Raman",
    title = "{The Scalar Chemical Potential in Cosmological Collider Physics}",
    eprint = "2010.04727",
    archivePrefix = "arXiv",
    primaryClass = "hep-ph",
    reportNumber = "UMD-PP-020-09",
    doi = "10.1007/JHEP02(2021)079",
    journal = "JHEP",
    volume = "02",
    pages = "079",
    year = "2021"
}

@ARTICLE{2009arXiv0912.0201L,
       author = {{LSST Science Collaboration} and {Abell}, Paul A. and {Allison}, Julius and {Anderson}, Scott F. and {Andrew}, John R. and {Angel}, J. Roger P. and {Armus}, Lee and {Arnett}, David and {Asztalos}, S.~J. and {Axelrod}, Tim S. and {Bailey}, Stephen and {Ballantyne}, D.~R. and {Bankert}, Justin R. and {Barkhouse}, Wayne A. and {Barr}, Jeffrey D. and {Barrientos}, L. Felipe and {Barth}, Aaron J. and {Bartlett}, James G. and {Becker}, Andrew C. and {Becla}, Jacek and {Beers}, Timothy C. and {Bernstein}, Joseph P. and {Biswas}, Rahul and {Blanton}, Michael R. and {Bloom}, Joshua S. and {Bochanski}, John J. and {Boeshaar}, Pat and {Borne}, Kirk D. and {Bradac}, Marusa and {Brandt}, W.~N. and {Bridge}, Carrie R. and {Brown}, Michael E. and {Brunner}, Robert J. and {Bullock}, James S. and {Burgasser}, Adam J. and {Burge}, James H. and {Burke}, David L. and {Cargile}, Phillip A. and {Chandrasekharan}, Srinivasan and {Chartas}, George and {Chesley}, Steven R. and {Chu}, You-Hua and {Cinabro}, David and {Claire}, Mark W. and {Claver}, Charles F. and {Clowe}, Douglas and {Connolly}, A.~J. and {Cook}, Kem H. and {Cooke}, Jeff and {Cooray}, Asantha and {Covey}, Kevin R. and {Culliton}, Christopher S. and {de Jong}, Roelof and {de Vries}, Willem H. and {Debattista}, Victor P. and {Delgado}, Francisco and {Dell'Antonio}, Ian P. and {Dhital}, Saurav and {Di Stefano}, Rosanne and {Dickinson}, Mark and {Dilday}, Benjamin and {Djorgovski}, S.~G. and {Dobler}, Gregory and {Donalek}, Ciro and {Dubois-Felsmann}, Gregory and {Durech}, Josef and {Eliasdottir}, Ardis and {Eracleous}, Michael and {Eyer}, Laurent and {Falco}, Emilio E. and {Fan}, Xiaohui and {Fassnacht}, Christopher D. and {Ferguson}, Harry C. and {Fernandez}, Yanga R. and {Fields}, Brian D. and {Finkbeiner}, Douglas and {Figueroa}, Eduardo E. and {Fox}, Derek B. and {Francke}, Harold and {Frank}, James S. and {Frieman}, Josh and {Fromenteau}, Sebastien and {Furqan}, Muhammad and {Galaz}, Gaspar and {Gal-Yam}, A. and {Garnavich}, Peter and {Gawiser}, Eric and {Geary}, John and {Gee}, Perry and {Gibson}, Robert R. and {Gilmore}, Kirk and {Grace}, Emily A. and {Green}, Richard F. and {Gressler}, William J. and {Grillmair}, Carl J. and {Habib}, Salman and {Haggerty}, J.~S. and {Hamuy}, Mario and {Harris}, Alan W. and {Hawley}, Suzanne L. and {Heavens}, Alan F. and {Hebb}, Leslie and {Henry}, Todd J. and {Hileman}, Edward and {Hilton}, Eric J. and {Hoadley}, Keri and {Holberg}, J.~B. and {Holman}, Matt J. and {Howell}, Steve B. and {Infante}, Leopoldo and {Ivezic}, Zeljko and {Jacoby}, Suzanne H. and {Jain}, Bhuvnesh and {R} and {Jedicke} and {Jee}, M. James and {Garrett Jernigan}, J. and {Jha}, Saurabh W. and {Johnston}, Kathryn V. and {Jones}, R. Lynne and {Juric}, Mario and {Kaasalainen}, Mikko and {Styliani} and {Kafka} and {Kahn}, Steven M. and {Kaib}, Nathan A. and {Kalirai}, Jason and {Kantor}, Jeff and {Kasliwal}, Mansi M. and {Keeton}, Charles R. and {Kessler}, Richard and {Knezevic}, Zoran and {Kowalski}, Adam and {Krabbendam}, Victor L. and {Krughoff}, K. Simon and {Kulkarni}, Shrinivas and {Kuhlman}, Stephen and {Lacy}, Mark and {Lepine}, Sebastien and {Liang}, Ming and {Lien}, Amy and {Lira}, Paulina and {Long}, Knox S. and {Lorenz}, Suzanne and {Lotz}, Jennifer M. and {Lupton}, R.~H. and {Lutz}, Julie and {Macri}, Lucas M. and {Mahabal}, Ashish A. and {Mandelbaum}, Rachel and {Marshall}, Phil and {May}, Morgan and {McGehee}, Peregrine M. and {Meadows}, Brian T. and {Meert}, Alan and {Milani}, Andrea and {Miller}, Christopher J. and {Miller}, Michelle and {Mills}, David and {Minniti}, Dante and {Monet}, David and {Mukadam}, Anjum S. and {Nakar}, Ehud and {Neill}, Douglas R. and {Newman}, Jeffrey A. and {Nikolaev}, Sergei and {Nordby}, Martin and {O'Connor}, Paul and {Oguri}, Masamune and {Oliver}, John and {Olivier}, Scot S. and {Olsen}, Julia K. and {Olsen}, Knut and {Olszewski}, Edward W. and {Oluseyi}, Hakeem and {Padilla}, Nelson D. and {Parker}, Alex and {Pepper}, Joshua and {Peterson}, John R. and {Petry}, Catherine and {Pinto}, Philip A. and {Pizagno}, James L. and {Popescu}, Bogdan and {Prsa}, Andrej and {Radcka}, Veljko and {Raddick}, M. Jordan and {Rasmussen}, Andrew and {Rau}, Arne and {Rho}, Jeonghee and {Rhoads}, James E. and {Richards}, Gordon T. and {Ridgway}, Stephen T. and {Robertson}, Brant E. and {Roskar}, Rok and {Saha}, Abhijit and {Sarajedini}, Ata and {Scannapieco}, Evan and {Schalk}, Terry and {Schindler}, Rafe and {Schmidt}, Samuel},
        title = "{LSST Science Book, Version 2.0}",
      journal = {arXiv e-prints},
         year = 2009,
        month = dec,
          eid = {arXiv:0912.0201},
        pages = {arXiv:0912.0201},
          doi = {10.48550/arXiv.0912.0201},
archivePrefix = {arXiv},
       eprint = {0912.0201},
 primaryClass = {astro-ph.IM},
       adsurl = {https://ui.adsabs.harvard.edu/abs/2009arXiv0912.0201L}
}

@article{McCulloch:2024hiz,
    author = "McCulloch, Ciaran and Pajer, Enrico and Tong, Xi",
    title = "{A cosmological tachyon collider: enhancing the long-short scale coupling}",
    eprint = "2401.11009",
    archivePrefix = "arXiv",
    primaryClass = "hep-th",
    doi = "10.1007/JHEP05(2024)262",
    journal = "JHEP",
    volume = "05",
    pages = "262",
    year = "2024"
}

@article{Chen:2009zp,
    author = "Chen, Xingang and Wang, Yi",
    title = "{Quasi-Single Field Inflation and Non-Gaussianities}",
    eprint = "0911.3380",
    archivePrefix = "arXiv",
    primaryClass = "hep-th",
    doi = "10.1088/1475-7516/2010/04/027",
    journal = "JCAP",
    volume = "04",
    pages = "027",
    year = "2010"
}

@ARTICLE{Angulo_2015,
       author = {{Angulo}, Raul E. and {Foreman}, Simon and {Schmittfull}, Marcel and {Senatore}, Leonardo},
        title = "{The one-loop matter bispectrum in the Effective Field Theory of Large Scale Structures}",
      journal = {\jcap},
         year = 2015,
        month = oct,
       volume = {2015},
       number = {10},
        pages = {039-039},
          doi = {10.1088/1475-7516/2015/10/039},
archivePrefix = {arXiv},
       eprint = {1406.4143},
 primaryClass = {astro-ph.CO},
       adsurl = {https://ui.adsabs.harvard.edu/abs/2015JCAP...10..039A}
}

@ARTICLE{Carrasco_2014,
       author = {{Carrasco}, John Joseph M. and {Foreman}, Simon and {Green}, Daniel and {Senatore}, Leonardo},
        title = "{The Effective Field Theory of Large Scale Structures at two loops}",
      journal = {\jcap},
         year = 2014,
        month = jul,
       volume = {2014},
       number = {7},
          eid = {057},
        pages = {057},
          doi = {10.1088/1475-7516/2014/07/057},
archivePrefix = {arXiv},
       eprint = {1310.0464},
 primaryClass = {astro-ph.CO},
       adsurl = {https://ui.adsabs.harvard.edu/abs/2014JCAP...07..057C}
}

@ARTICLE{Vlah_2015,
       author = {{Vlah}, Zvonimir and {White}, Martin and {Aviles}, Alejandro},
        title = "{A Lagrangian effective field theory}",
      journal = {\jcap},
         year = 2015,
        month = sep,
       volume = {2015},
       number = {9},
        pages = {014-014},
          doi = {10.1088/1475-7516/2015/09/014},
archivePrefix = {arXiv},
       eprint = {1506.05264},
 primaryClass = {astro-ph.CO},
       adsurl = {https://ui.adsabs.harvard.edu/abs/2015JCAP...09..014V}
}

@ARTICLE{Senatore_2014,
       author = {{Senatore}, Leonardo and {Zaldarriaga}, Matias},
        title = "{Redshift Space Distortions in the Effective Field Theory of Large Scale Structures}",
      journal = {arXiv e-prints},
         year = 2014,
        month = sep,
          eid = {arXiv:1409.1225},
        pages = {arXiv:1409.1225},
          doi = {10.48550/arXiv.1409.1225},
archivePrefix = {arXiv},
       eprint = {1409.1225},
 primaryClass = {astro-ph.CO},
       adsurl = {https://ui.adsabs.harvard.edu/abs/2014arXiv1409.1225S}
}

@ARTICLE{Senatore_2015,
       author = {{Senatore}, Leonardo and {Zaldarriaga}, Matias},
        title = "{The IR-resummed Effective Field Theory of Large Scale Structures}",
      journal = {\jcap},
         year = 2015,
        month = feb,
       volume = {2015},
       number = {2},
        pages = {013-013},
          doi = {10.1088/1475-7516/2015/02/013},
archivePrefix = {arXiv},
       eprint = {1404.5954},
 primaryClass = {astro-ph.CO},
       adsurl = {https://ui.adsabs.harvard.edu/abs/2015JCAP...02..013S}
}

@ARTICLE{Carrasco_2012,
       author = {{Carrasco}, John Joseph M. and {Hertzberg}, Mark P. and {Senatore}, Leonardo},
        title = "{The effective field theory of cosmological large scale structures}",
      journal = {Journal of High Energy Physics},
         year = 2012,
        month = sep,
       volume = {2012},
          eid = {82},
        pages = {82},
          doi = {10.1007/JHEP09(2012)082},
archivePrefix = {arXiv},
       eprint = {1206.2926},
 primaryClass = {astro-ph.CO},
       adsurl = {https://ui.adsabs.harvard.edu/abs/2012JHEP...09..082C}
}

@ARTICLE{Baumann_2012,
       author = {{Baumann}, Daniel and {Nicolis}, Alberto and {Senatore}, Leonardo and {Zaldarriaga}, Matias},
        title = "{Cosmological non-linearities as an effective fluid}",
      journal = {\jcap},
         year = 2012,
        month = jul,
       volume = {2012},
       number = {7},
          eid = {051},
        pages = {051},
          doi = {10.1088/1475-7516/2012/07/051},
archivePrefix = {arXiv},
       eprint = {1004.2488},
 primaryClass = {astro-ph.CO},
       adsurl = {https://ui.adsabs.harvard.edu/abs/2012JCAP...07..051B}
}

@ARTICLE{Baldauf_2021,
       author = {{Baldauf}, Tobias and {Garny}, Mathias and {Taule}, Petter and {Steele}, Theo},
        title = "{Two-loop bispectrum of large-scale structure}",
      journal = {\prd},
         year = 2021,
        month = dec,
       volume = {104},
       number = {12},
          eid = {123551},
        pages = {123551},
          doi = {10.1103/PhysRevD.104.123551},
archivePrefix = {arXiv},
       eprint = {2110.13930},
 primaryClass = {astro-ph.CO},
       adsurl = {https://ui.adsabs.harvard.edu/abs/2021PhRvD.104l3551B}
}

@ARTICLE{Bakx_2026,
       author = {{Bakx}, Thomas and {Rubira}, Henrique and {Chisari}, Nora Elisa and {Vlah}, Zvonimir},
        title = "{Rapid cosmological inference with the two-loop matter power spectrum}",
      journal = {The Open Journal of Astrophysics},
         year = 2026,
        month = feb,
       volume = {9},
        pages = {57501},
          doi = {10.33232/001c.157501},
archivePrefix = {arXiv},
       eprint = {2508.00611},
 primaryClass = {astro-ph.CO},
       adsurl = {https://ui.adsabs.harvard.edu/abs/2026OJAp....957501B}
}

@ARTICLE{Planck_2020_NG,
       author = {{Planck Collaboration} and {Akrami}, Y. and {Arroja}, F. and {Ashdown}, M. and {Aumont}, J. and {Baccigalupi}, C. and {Ballardini}, M. and {Banday}, A.~J. and {Barreiro}, R.~B. and {Bartolo}, N. and {Basak}, S. and {Benabed}, K. and {Bernard}, J.-P. and {Bersanelli}, M. and {Bielewicz}, P. and {Bond}, J.~R. and {Borrill}, J. and {Bouchet}, F.~R. and {Bucher}, M. and {Burigana}, C. and {Butler}, R.~C. and {Calabrese}, E. and {Cardoso}, J.-F. and {Casaponsa}, B. and {Challinor}, A. and {Chiang}, H.~C. and {Colombo}, L.~P.~L. and {Combet}, C. and {Crill}, B.~P. and {Cuttaia}, F. and {de Bernardis}, P. and {de Rosa}, A. and {de Zotti}, G. and {Delabrouille}, J. and {Delouis}, J.-M. and {Di Valentino}, E. and {Diego}, J.~M. and {Dor{\'e}}, O. and {Douspis}, M. and {Ducout}, A. and {Dupac}, X. and {Dusini}, S. and {Efstathiou}, G. and {Elsner}, F. and {En{\ss}lin}, T.~A. and {Eriksen}, H.~K. and {Fantaye}, Y. and {Fergusson}, J. and {Fernandez-Cobos}, R. and {Finelli}, F. and {Frailis}, M. and {Fraisse}, A.~A. and {Franceschi}, E. and {Frolov}, A. and {Galeotta}, S. and {Galli}, S. and {Ganga}, K. and {G{\'e}nova-Santos}, R.~T. and {Gerbino}, M. and {Gonz{\'a}lez-Nuevo}, J. and {G{\'o}rski}, K.~M. and {Gratton}, S. and {Gruppuso}, A. and {Gudmundsson}, J.~E. and {Hamann}, J. and {Handley}, W. and {Hansen}, F.~K. and {Herranz}, D. and {Hivon}, E. and {Huang}, Z. and {Jaffe}, A.~H. and {Jones}, W.~C. and {Jung}, G. and {Keih{\"a}nen}, E. and {Keskitalo}, R. and {Kiiveri}, K. and {Kim}, J. and {Krachmalnicoff}, N. and {Kunz}, M. and {Kurki-Suonio}, H. and {Lamarre}, J.-M. and {Lasenby}, A. and {Lattanzi}, M. and {Lawrence}, C.~R. and {Le Jeune}, M. and {Levrier}, F. and {Lewis}, A. and {Liguori}, M. and {Lilje}, P.~B. and {Lindholm}, V. and {L{\'o}pez-Caniego}, M. and {Ma}, Y.-Z. and {Mac{\'\i}as-P{\'e}rez}, J.~F. and {Maggio}, G. and {Maino}, D. and {Mandolesi}, N. and {Marcos-Caballero}, A. and {Maris}, M. and {Martin}, P.~G. and {Mart{\'\i}nez-Gonz{\'a}lez}, E. and {Matarrese}, S. and {Mauri}, N. and {McEwen}, J.~D. and {Meerburg}, P.~D. and {Meinhold}, P.~R. and {Melchiorri}, A. and {Mennella}, A. and {Migliaccio}, M. and {Miville-Desch{\^e}nes}, M.-A. and {Molinari}, D. and {Moneti}, A. and {Montier}, L. and {Morgante}, G. and {Moss}, A. and {M{\"u}nchmeyer}, M. and {Natoli}, P. and {Oppizzi}, F. and {Pagano}, L. and {Paoletti}, D. and {Partridge}, B. and {Patanchon}, G. and {Perrotta}, F. and {Pettorino}, V. and {Piacentini}, F. and {Polenta}, G. and {Puget}, J.-L. and {Rachen}, J.~P. and {Racine}, B. and {Reinecke}, M. and {Remazeilles}, M. and {Renzi}, A. and {Rocha}, G. and {Rubi{\~n}o-Mart{\'\i}n}, J.~A. and {Ruiz-Granados}, B. and {Salvati}, L. and {Savelainen}, M. and {Scott}, D. and {Shellard}, E.~P.~S. and {Shiraishi}, M. and {Sirignano}, C. and {Sirri}, G. and {Smith}, K. and {Spencer}, L.~D. and {Stanco}, L. and {Sunyaev}, R. and {Suur-Uski}, A.-S. and {Tauber}, J.~A. and {Tavagnacco}, D. and {Tenti}, M. and {Toffolatti}, L. and {Tomasi}, M. and {Trombetti}, T. and {Valiviita}, J. and {Van Tent}, B. and {Vielva}, P. and {Villa}, F. and {Vittorio}, N. and {Wandelt}, B.~D. and {Wehus}, I.~K. and {Zacchei}, A. and {Zonca}, A.},
        title = "{Planck 2018 results. IX. Constraints on primordial non-Gaussianity}",
      journal = {\aap},
         year = 2020,
        month = sep,
       volume = {641},
          eid = {A9},
        pages = {A9},
          doi = {10.1051/0004-6361/201935891},
archivePrefix = {arXiv},
       eprint = {1905.05697},
 primaryClass = {astro-ph.CO},
       adsurl = {https://ui.adsabs.harvard.edu/abs/2020A&A...641A...9P}
}

@article{Komatsu:2001rj,
    author = "Komatsu, Eiichiro and Spergel, David N.",
    title = "{Acoustic signatures in the primary microwave background bispectrum}",
    eprint = "astro-ph/0005036",
    archivePrefix = "arXiv",
    doi = "10.1103/PhysRevD.63.063002",
    journal = "Phys. Rev. D",
    volume = "63",
    pages = "063002",
    year = "2001"
}

@article{Dalal:2007cu,
    author = "Dalal, Neal and Dore, Olivier and Huterer, Dragan and Shirokov, Alexander",
    title = "{The imprints of primordial non-gaussianities on large-scale structure: scale dependent bias and abundance of virialized objects}",
    eprint = "0710.4560",
    archivePrefix = "arXiv",
    primaryClass = "astro-ph",
    doi = "10.1103/PhysRevD.77.123514",
    journal = "Phys. Rev. D",
    volume = "77",
    pages = "123514",
    year = "2008"
}

@article{Matarrese:2008nc,
    author = "Matarrese, Sabino and Verde, Licia",
    title = "{The effect of primordial non-Gaussianity on halo bias}",
    eprint = "0801.4826",
    archivePrefix = "arXiv",
    primaryClass = "astro-ph",
    doi = "10.1086/587840",
    journal = "Astrophys. J. Lett.",
    volume = "677",
    pages = "L77--L80",
    year = "2008"
}

@article{Slosar:2008hx,
    author = "Slosar, Anze and Hirata, Christopher and Seljak, Uros and Ho, Shirley and Padmanabhan, Nikhil",
    title = "{Constraints on local primordial non-Gaussianity from large scale structure}",
    eprint = "0805.3580",
    archivePrefix = "arXiv",
    primaryClass = "astro-ph",
    doi = "10.1088/1475-7516/2008/08/031",
    journal = "JCAP",
    volume = "08",
    pages = "031",
    year = "2008"
}

@article{Schmidt:2010gw,
    author = "Schmidt, Fabian and Kamionkowski, Marc",
    title = "{Halo Clustering with Non-Local Non-Gaussianity}",
    eprint = "1008.0638",
    archivePrefix = "arXiv",
    primaryClass = "astro-ph.CO",
    doi = "10.1103/PhysRevD.82.103002",
    journal = "Phys. Rev. D",
    volume = "82",
    pages = "103002",
    year = "2010"
}

@article{Assassi:2015fma,
    author = "Assassi, Valentin and Baumann, Daniel and Schmidt, Fabian",
    title = "{Galaxy Bias and Primordial Non-Gaussianity}",
    eprint = "1510.03723",
    archivePrefix = "arXiv",
    primaryClass = "astro-ph.CO",
    doi = "10.1088/1475-7516/2015/12/043",
    journal = "JCAP",
    volume = "12",
    pages = "043",
    year = "2015"
}

@article{MoradinezhadDizgah:2017szk,
    author = "Moradinezhad Dizgah, Azadeh and Dvorkin, Cora",
    title = "{Scale-Dependent Galaxy Bias from Massive Particles with Spin during Inflation}",
    eprint = "1708.06473",
    archivePrefix = "arXiv",
    primaryClass = "astro-ph.CO",
    doi = "10.1088/1475-7516/2018/01/010",
    journal = "JCAP",
    volume = "01",
    pages = "010",
    year = "2018"
}

@article{Arkani-Hamed:2015bza,
    author = "Arkani-Hamed, Nima and Maldacena, Juan",
    title = "{Cosmological Collider Physics}",
    eprint = "1503.08043",
    archivePrefix = "arXiv",
    primaryClass = "hep-th",
    journal = {}
}

@article{Lee:2016vti,
    author = "Lee, Hayden and Baumann, Daniel and Pimentel, Guilherme L.",
    title = "{Non-Gaussianity as a Particle Detector}",
    eprint = "1607.03735",
    archivePrefix = "arXiv",
    primaryClass = "hep-th",
    doi = "10.1007/JHEP12(2016)040",
    journal = "JHEP",
    volume = "12",
    pages = "040",
    year = "2016"
}

@article{Bordin:2018pca,
    author = "Bordin, Lorenzo and Creminelli, Paolo and Khmelnitsky, Andrei and Senatore, Leonardo",
    title = "{Light Particles with Spin in Inflation}",
    eprint = "1806.10587",
    archivePrefix = "arXiv",
    primaryClass = "hep-th",
    doi = "10.1088/1475-7516/2018/10/013",
    journal = "JCAP",
    volume = "10",
    pages = "013",
    year = "2018"
}

@article{Agullo:2012cs,
    author = "Agullo, Ivan and Shandera, Sarah",
    title = "{Large non-Gaussian Halo Bias from Single Field Inflation}",
    eprint = "1204.4409",
    archivePrefix = "arXiv",
    primaryClass = "astro-ph.CO",
    doi = "10.1088/1475-7516/2012/09/007",
    journal = "JCAP",
    volume = "09",
    pages = "007",
    year = "2012"
}

@article{Mirbabayi:2022cbt,
    author = "Mirbabayi, Mehrdad and Gruzinov, Andrei",
    title = "{Shapes of non-Gaussianity in warm inflation}",
    eprint = "2205.13227",
    archivePrefix = "arXiv",
    primaryClass = "astro-ph.CO",
    doi = "10.1088/1475-7516/2023/02/012",
    journal = "JCAP",
    volume = "02",
    pages = "012",
    year = "2023"
}

@article{Endlich:2012pz,
    author = "Endlich, Solomon and Nicolis, Alberto and Wang, Junpu",
    title = "{Solid Inflation}",
    eprint = "1210.0569",
    archivePrefix = "arXiv",
    primaryClass = "hep-th",
    doi = "10.1088/1475-7516/2013/10/011",
    journal = "JCAP",
    volume = "10",
    pages = "011",
    year = "2013"
}

@article{Schmidt:2012ky,
    author = "Schmidt, Fabian and Hui, Lam",
    title = "{Cosmic Microwave Background Power Asymmetry from Non-Gaussian Modulation}",
    eprint = "1210.2965",
    archivePrefix = "arXiv",
    primaryClass = "astro-ph.CO",
    doi = "10.1103/PhysRevLett.110.011301",
    journal = "Phys. Rev. Lett.",
    volume = "110",
    pages = "011301",
    year = "2013",
    note = "[Erratum: Phys.Rev.Lett. 110, 059902 (2013)]"
}

@article{Shiraishi:2013vja,
    author = "Shiraishi, Maresuke and Komatsu, Eiichiro and Peloso, Marco and Barnaby, Neil",
    title = "{Signatures of anisotropic sources in the squeezed-limit bispectrum of the cosmic microwave background}",
    eprint = "1302.3056",
    archivePrefix = "arXiv",
    primaryClass = "astro-ph.CO",
    doi = "10.1088/1475-7516/2013/05/002",
    journal = "JCAP",
    volume = "05",
    pages = "002",
    year = "2013"
}

@article{Assassi:2015jqa,
    author = "Assassi, Valentin and Baumann, Daniel and Pajer, Enrico and Welling, Yvette and van der Woude, Drian",
    title = "{Effective theory of large-scale structure with primordial non-Gaussianity}",
    eprint = "1505.06668",
    archivePrefix = "arXiv",
    primaryClass = "astro-ph.CO",
    doi = "10.1088/1475-7516/2015/11/024",
    journal = "JCAP",
    volume = "11",
    pages = "024",
    year = "2015"
}

@article{Schmidt:2015xka,
    author = "Schmidt, Fabian and Chisari, Nora Elisa and Dvorkin, Cora",
    title = "{Imprint of inflation on galaxy shape correlations}",
    eprint = "1506.02671",
    archivePrefix = "arXiv",
    primaryClass = "astro-ph.CO",
    doi = "10.1088/1475-7516/2015/10/032",
    journal = "JCAP",
    volume = "10",
    pages = "032",
    year = "2015"
}

@ARTICLE{Akitsu-2021,
       author = {{Akitsu}, Kazuyuki and {Kurita}, Toshiki and {Nishimichi}, Takahiro and {Takada}, Masahiro and {Tanaka}, Satoshi},
        title = "{Imprint of anisotropic primordial non-Gaussianity on halo intrinsic alignments in simulations}",
      journal = {\prd},
         year = 2021,
        month = apr,
       volume = {103},
       number = {8},
          eid = {083508},
        pages = {083508},
          doi = {10.1103/PhysRevD.103.083508},
archivePrefix = {arXiv},
       eprint = {2007.03670},
 primaryClass = {astro-ph.CO},
       adsurl = {https://ui.adsabs.harvard.edu/abs/2021PhRvD.103h3508A}
}

@article{Mirbabayi:2014zca,
    author = "Mirbabayi, Mehrdad and Schmidt, Fabian and Zaldarriaga, Matias",
    title = "{Biased Tracers and Time Evolution}",
    eprint = "1412.5169",
    archivePrefix = "arXiv",
    primaryClass = "astro-ph.CO",
    doi = "10.1088/1475-7516/2015/07/030",
    journal = "JCAP",
    volume = "07",
    pages = "030",
    year = "2015"
}

@article{Desjacques:2018pfv,
    author = "Desjacques, Vincent and Jeong, Donghui and Schmidt, Fabian",
    title = "{The Galaxy Power Spectrum and Bispectrum in Redshift Space}",
    eprint = "1806.04015",
    archivePrefix = "arXiv",
    primaryClass = "astro-ph.CO",
    doi = "10.1088/1475-7516/2018/12/035",
    journal = "JCAP",
    volume = "12",
    pages = "035",
    year = "2018"
}

@ARTICLE{Obuljen-2019,
       author = {{Obuljen}, Andrej and {Dalal}, Neal and {Percival}, Will J.},
        title = "{Anisotropic halo assembly bias and redshift-space distortions}",
      journal = {\jcap},
         year = 2019,
        month = oct,
       volume = {2019},
       number = {10},
          eid = {020},
        pages = {020},
          doi = {10.1088/1475-7516/2019/10/020},
archivePrefix = {arXiv},
       eprint = {1906.11823},
 primaryClass = {astro-ph.CO},
       adsurl = {https://ui.adsabs.harvard.edu/abs/2019JCAP...10..020O}
}

@ARTICLE{Croft-1998,
       author = {{Croft}, Rupert A.~C. and {Weinberg}, David H. and {Katz}, Neal and {Hernquist}, Lars},
        title = "{Recovery of the Power Spectrum of Mass Fluctuations from Observations of the Ly{\ensuremath{\alpha}} Forest}",
      journal = {\apj},
         year = 1998,
        month = mar,
       volume = {495},
       number = {1},
        pages = {44-62},
          doi = {10.1086/305289},
archivePrefix = {arXiv},
       eprint = {astro-ph/9708018},
 primaryClass = {astro-ph},
       adsurl = {https://ui.adsabs.harvard.edu/abs/1998ApJ...495...44C}
}

@ARTICLE{Ivanov-2023,
       author = {{Ivanov}, Mikhail M. and {Philcox}, Oliver H.~E. and {Cabass}, Giovanni and {Nishimichi}, Takahiro and {Simonovi{\'c}}, Marko and {Zaldarriaga}, Matias},
        title = "{Cosmology with the galaxy bispectrum multipoles: Optimal estimation and application to BOSS data}",
      journal = {\prd},
         year = 2023,
        month = apr,
       volume = {107},
       number = {8},
          eid = {083515},
        pages = {083515},
          doi = {10.1103/PhysRevD.107.083515},
archivePrefix = {arXiv},
       eprint = {2302.04414},
 primaryClass = {astro-ph.CO},
       adsurl = {https://ui.adsabs.harvard.edu/abs/2023PhRvD.107h3515I}
}

@ARTICLE{Bock_2026,
       author = {{Bock}, James J. and {Aboobaker}, Asad M. and {Adamo}, Joseph and {Akeson}, Rachel and {Alred}, John M. and {Alibay}, Farah and {Ashby}, Matthew L.~N. and {Bach}, Yoonsoo P. and {Bleem}, Lindsey E. and {Bolton}, Douglas and {Braun}, David F. and {Bruton}, Sean and {Bryan}, Sean A. and {Chang}, Tzu-Ching and {Chen}, Shuang-Shuang and {Cheng}, Yun-Ting and {Cheshire}, IV, James R. and {Chiang}, Yi-Kuan and {Choppin de Janvry}, Jean and {Condon}, Samuel and {Cook}, Walter R. and {Cooray}, Asantha and {Crill}, Brendan P. and {Cukierman}, Ari J. and {Dor{\'e}}, Olivier and {Dowell}, C. Darren and {Dubois-Felsmann}, Gregory P. and {Eifler}, Tim and {Everett}, Spencer and {Fabinsky}, Beth E. and {Faisst}, Andreas L. and {Fanson}, James L. and {Farrington}, Allen H. and {Fatahi}, Tamim and {Fazar}, Candice M. and {Feder}, Richard M. and {Frater}, Eric H. and {Grasshorn Gebhardt}, Henry S. and {Giri}, Utkarsh and {Goldina}, Tatiana and {Gorjian}, Varoujan and {Habib}, Salman and {Hart}, William G. and {Heinrich}, Chen and {Hora}, Joseph L. and {Huai}, Zhaoyu and {Hui}, Howard and {Jo}, Young-Soo and {Jeong}, Woong-Seob and {Kang}, Jae Hwan and {Kang}, Miju and {Kecman}, Branislav and {Kim}, Chul-Hwan and {Kim}, Jaeyeong and {Kim}, Minjin and {Kim}, Young-Jun and {Kim}, Yongjung and {Kirkpatrick}, J. Davy and {Kobayashi}, Yosuke and {Korngut}, Phil M. and {Krause}, Elisabeth and {Lee}, Bomee and {Lee}, Ho-Gyu and {Lee}, Jae-Joon and {Lee}, Jeong-Eun and {Lisse}, Carey M. and {Mariani}, Giacomo and {Masters}, Daniel C. and {Mauskopf}, Philip D. and {Melnick}, Gary J. and {Minasyan}, Mary H. and {Mirocha}, Jordan and {Miyasaka}, Hiromasa and {Moore}, Anne and {Moore}, Bradley D. and {Murgia}, Giulia and {Naylor}, Bret J. and {Nelson}, Christina and {Nguyen}, Chi H. and {Nguyen}, Hien T. and {Noh}, Jinyoung K. and {Padin}, Stephen and {Paladini}, Roberta and {Park}, Sung-Joon and {Penanen}, Konstantin I. and {Putnam}, Dustin S. and {Pyo}, Jeonghyun and {Ramachandra}, Nesar and {Ramanathan}, Keshav and {Rustamkulov}, Zafar and {Reiley}, Daniel J. and {Rice}, Eric B. and {Rocca}, Jennifer M. and {Seok}, Ji Yeon and {Smith}, Roger and {Stober}, Jeremy and {Susca}, Sara and {Teplitz}, Harry I. and {Thelen}, Michael P. and {Tolls}, Volker and {Torrini}, Gabriela and {Trangsrud}, Amy R. and {Unwin}, Stephen and {Velicheti}, Phani and {Wang}, Pao-Yu and {Wen}, Robin Y. and {Werner}, Michael W. and {Williams}, Abby E. and {Williamson}, Ross and {Wincentsen}, James and {Windhorst}, Rogier A. and {Yang}, Soung-Chul and {Yang}, Yujin and {Zemcov}, Michael},
        title = "{The SPHEREx Satellite Mission}",
      journal = {\apj},
         year = 2026,
        month = mar,
       volume = {999},
       number = {1},
          eid = {139},
        pages = {139},
          doi = {10.3847/1538-4357/ae2be2},
archivePrefix = {arXiv},
       eprint = {2511.02985},
 primaryClass = {astro-ph.IM},
       adsurl = {https://ui.adsabs.harvard.edu/abs/2026ApJ...999..139B}
}

@ARTICLE{SO_2019,
       author = {{Ade}, Peter and {Aguirre}, James and {Ahmed}, Zeeshan and {Aiola}, Simone and {Ali}, Aamir and {Alonso}, David and {Alvarez}, Marcelo A. and {Arnold}, Kam and {Ashton}, Peter and {Austermann}, Jason and {Awan}, Humna and {Baccigalupi}, Carlo and {Baildon}, Taylor and {Barron}, Darcy and {Battaglia}, Nick and {Battye}, Richard and {Baxter}, Eric and {Bazarko}, Andrew and {Beall}, James A. and {Bean}, Rachel and {Beck}, Dominic and {Beckman}, Shawn and {Beringue}, Benjamin and {Bianchini}, Federico and {Boada}, Steven and {Boettger}, David and {Bond}, J. Richard and {Borrill}, Julian and {Brown}, Michael L. and {Bruno}, Sarah Marie and {Bryan}, Sean and {Calabrese}, Erminia and {Calafut}, Victoria and {Calisse}, Paolo and {Carron}, Julien and {Challinor}, Anthony and {Chesmore}, Grace and {Chinone}, Yuji and {Chluba}, Jens and {Cho}, Hsiao-Mei Sherry and {Choi}, Steve and {Coppi}, Gabriele and {Cothard}, Nicholas F. and {Coughlin}, Kevin and {Crichton}, Devin and {Crowley}, Kevin D. and {Crowley}, Kevin T. and {Cukierman}, Ari and {D'Ewart}, John M. and {D{\"u}nner}, Rolando and {de Haan}, Tijmen and {Devlin}, Mark and {Dicker}, Simon and {Didier}, Joy and {Dobbs}, Matt and {Dober}, Bradley and {Duell}, Cody J. and {Duff}, Shannon and {Duivenvoorden}, Adri and {Dunkley}, Jo and {Dusatko}, John and {Errard}, Josquin and {Fabbian}, Giulio and {Feeney}, Stephen and {Ferraro}, Simone and {Flux{\`a}}, Pedro and {Freese}, Katherine and {Frisch}, Josef C. and {Frolov}, Andrei and {Fuller}, George and {Fuzia}, Brittany and {Galitzki}, Nicholas and {Gallardo}, Patricio A. and {Tomas Galvez Ghersi}, Jose and {Gao}, Jiansong and {Gawiser}, Eric and {Gerbino}, Martina and {Gluscevic}, Vera and {Goeckner-Wald}, Neil and {Golec}, Joseph and {Gordon}, Sam and {Gralla}, Megan and {Green}, Daniel and {Grigorian}, Arpi and {Groh}, John and {Groppi}, Chris and {Guan}, Yilun and {Gudmundsson}, Jon E. and {Han}, Dongwon and {Hargrave}, Peter and {Hasegawa}, Masaya and {Hasselfield}, Matthew and {Hattori}, Makoto and {Haynes}, Victor and {Hazumi}, Masashi and {He}, Yizhou and {Healy}, Erin and {Henderson}, Shawn W. and {Hervias-Caimapo}, Carlos and {Hill}, Charles A. and {Hill}, J. Colin and {Hilton}, Gene and {Hilton}, Matt and {Hincks}, Adam D. and {Hinshaw}, Gary and {Hlo{\v{z}}ek}, Ren{\'e}e and {Ho}, Shirley and {Ho}, Shuay-Pwu Patty and {Howe}, Logan and {Huang}, Zhiqi and {Hubmayr}, Johannes and {Huffenberger}, Kevin and {Hughes}, John P. and {Ijjas}, Anna and {Ikape}, Margaret and {Irwin}, Kent and {Jaffe}, Andrew H. and {Jain}, Bhuvnesh and {Jeong}, Oliver and {Kaneko}, Daisuke and {Karpel}, Ethan D. and {Katayama}, Nobuhiko and {Keating}, Brian and {Kernasovskiy}, Sarah S. and {Keskitalo}, Reijo and {Kisner}, Theodore and {Kiuchi}, Kenji and {Klein}, Jeff and {Knowles}, Kenda and {Koopman}, Brian and {Kosowsky}, Arthur and {Krachmalnicoff}, Nicoletta and {Kuenstner}, Stephen E. and {Kuo}, Chao-Lin and {Kusaka}, Akito and {Lashner}, Jacob and {Lee}, Adrian and {Lee}, Eunseong and {Leon}, David and {Leung}, Jason S.-Y. and {Lewis}, Antony and {Li}, Yaqiong and {Li}, Zack and {Limon}, Michele and {Linder}, Eric and {Lopez-Caraballo}, Carlos and {Louis}, Thibaut and {Lowry}, Lindsay and {Lungu}, Marius and {Madhavacheril}, Mathew and {Mak}, Daisy and {Maldonado}, Felipe and {Mani}, Hamdi and {Mates}, Ben and {Matsuda}, Frederick and {Maurin}, Lo{\"\i}c and {Mauskopf}, Phil and {May}, Andrew and {McCallum}, Nialh and {McKenney}, Chris and {McMahon}, Jeff and {Meerburg}, P. Daniel and {Meyers}, Joel and {Miller}, Amber and {Mirmelstein}, Mark and {Moodley}, Kavilan and {Munchmeyer}, Moritz and {Munson}, Charles and {Naess}, Sigurd and {Nati}, Federico and {Navaroli}, Martin and {Newburgh}, Laura and {Nguyen}, Ho Nam and {Niemack}, Michael and {Nishino}, Haruki and {Orlowski-Scherer}, John and {Page}, Lyman and {Partridge}, Bruce and {Peloton}, Julien and {Perrotta}, Francesca and {Piccirillo}, Lucio and {Pisano}, Giampaolo and {Poletti}, Davide and {Puddu}, Roberto and {Puglisi}, Giuseppe and {Raum}, Chris and {Reichardt}, Christian L. and {Remazeilles}, Mathieu and {Rephaeli}, Yoel and {Riechers}, Dominik and {Rojas}, Felipe and {Roy}, Anirban and {Sadeh}, Sharon and {Sakurai}, Yuki and {Salatino}, Maria and {Sathyanarayana Rao}, Mayuri and {Schaan}, Emmanuel and {Schmittfull}, Marcel and {Sehgal}, Neelima and {Seibert}, Joseph},
        title = "{The Simons Observatory: science goals and forecasts}",
      journal = {\jcap},
         year = 2019,
        month = feb,
       volume = {2019},
       number = {2},
          eid = {056},
        pages = {056},
          doi = {10.1088/1475-7516/2019/02/056},
archivePrefix = {arXiv},
       eprint = {1808.07445},
 primaryClass = {astro-ph.CO},
       adsurl = {https://ui.adsabs.harvard.edu/abs/2019JCAP...02..056A}
}

@ARTICLE{Schlegel_2022,
       author = {{Schlegel}, David J. and {Ferraro}, Simone and {Aldering}, Greg and {Baltay}, Charles and {BenZvi}, Segev and {Besuner}, Robert and {Blanc}, Guillermo A. and {Bolton}, Adam S. and {Bonaca}, Ana and {Brooks}, David and {Buckley-Geer}, Elizabeth and {Cai}, Zheng and {DeRose}, Joseph and {Dey}, Arjun and {Doel}, Peter and {Drlica-Wagner}, Alex and {Fan}, Xiaohui and {Gutierrez}, Gaston and {Green}, Daniel and {Guy}, Julien and {Huterer}, Dragan and {Infante}, Leopoldo and {Jelinsky}, Patrick and {Karagiannis}, Dionysios and {Kent}, Stephen M. and {Kim}, Alex G. and {Kneib}, Jean-Paul and {Kollmeier}, Juna A. and {Kremin}, Anthony and {Lahav}, Ofer and {Landriau}, Martin and {Lang}, Dustin and {Leauthaud}, Alexie and {Levi}, Michael E. and {Linder}, Eric V. and {Magneville}, Christophe and {Martini}, Paul and {McDonald}, Patrick and {Miller}, Christopher J. and {Myers}, Adam D. and {Newman}, Jeffrey A. and {Nugent}, Peter E. and {Palanque-Delabrouille}, Nathalie and {Padmanabhan}, Nikhil and {Palmese}, Antonella and {Poppett}, Claire and {Prochaska}, Jason X. and {Raichoor}, Anand and {Ramirez}, Solange and {Sailer}, Noah and {Schaan}, Emmanuel and {Schubnell}, Michael and {Seljak}, Uros and {Seo}, Hee-Jong and {Silber}, Joseph and {Simon}, Joshua D. and {Slepian}, Zachary and {Soares-Santos}, Marcelle and {Tarle}, Greg and {Valluri}, Monica and {Weaverdyck}, Noah J. and {Wechsler}, Risa H. and {White}, Martin and {Yeche}, Christophe and {Zhou}, Rongpu},
        title = "{A Spectroscopic Road Map for Cosmic Frontier: DESI, DESI-II, Stage-5}",
      journal = {arXiv e-prints},
         year = 2022,
        month = sep,
          eid = {arXiv:2209.03585},
        pages = {arXiv:2209.03585},
          doi = {10.48550/arXiv.2209.03585},
archivePrefix = {arXiv},
       eprint = {2209.03585},
 primaryClass = {astro-ph.CO},
       adsurl = {https://ui.adsabs.harvard.edu/abs/2022arXiv220903585S}
}

@ARTICLE{Besuner_2025,
       author = {{Besuner}, Robert and {Dey}, Arjun and {Drlica-Wagner}, Alex and {Ebina}, Haruki and {Fernandez Moroni}, Guillermo and {Ferraro}, Simone and {Forero-Romero}, Jaime and {Honscheid}, Klaus and {Jelinsky}, Pat and {Lang}, Dustin and {Levi}, Michael and {Martini}, Paul and {Myers}, Adam and {Palanque-Delabrouille}, Nathalie and {Panda}, Swayamtrupta and {Poppett}, Claire and {Sailer}, Noah and {Schlegel}, David and {Shafieloo}, Arman and {Silber}, Joseph and {White}, Martin and {Abbott}, Timothy and {Allen}, Lori and {Avila}, Santiago and {Avil{\'e}s}, Roberto and {Bailey}, Stephen and {Bault}, Abby and {Bouri}, Mohamed and {Boutsia}, Konstantina and {Burtin}, Eienne and {Chierchie}, Fernando and {Coulton}, William and {Dawson}, Kyle and {Dey}, Biprateep and {Dor{\'e}}, Olivier and {Dunlop}, Patrick and {Eisenstein}, Daniel and {Emanuele}, Castorina and {Escoffier}, Stephanie and {Estrada}, Juan and {Fagrelius}, Parker and {Fanning}, Kevin and {Fanning}, Timothy and {Font-Ribera}, Andreu and {Frieman}, Joshua and {Galal}, Malak and {Gluscevic}, Vera and {Gontcho}, Satya Gontcho A and {Green}, Daniel and {Gutierrez}, Gaston and {Guy}, Julien and {Hashemi}, Kevan and {Heathcote}, Steve and {Holland}, Steve and {Hou}, Jiamin and {Huterer}, Dragan and {Irigoyen Gimenez}, Blas and {Ivanov}, Mikhail and {Joyce}, Richard and {Jullo}, Eric and {Juneau}, Stephanie and {Juramy}, Claire and {Karcher}, Armin and {Kent}, Stephen and {Kirkby}, David and {Kneib}, Jean-Paul and {Krause}, Elisabeth and {Krolewski}, Alex and {Lahav}, Ofer and {Lapi}, Agustin and {Leauthaud}, Alexie and {Lewandowski}, Matthew and {Li}, Ting and {Lin}, Kenneth and {Loverde}, Marilena and {MacBride}, Sean and {Magneville}, Christophe and {Marshall}, Jennifer and {McDonald}, Patrick and {Miller}, Timothy and {Moustakas}, John and {M{\"u}nchmeyer}, Moritz and {Najita}, Joan and {Newman}, Jeff and {Percival}, Will and {Philcox}, Oliver and {Pires}, Priscila and {Raichoor}, Anand and {Roach}, Brandon and {Rockosi}, Constance and {Rombach}, Maxime and {Ross}, Ashley and {Sanchez}, Eusebio and {Schmidt}, Luke and {Schubnell}, Michael and {Sebok}, Rebekah and {Seljak}, Uros and {Silverstein}, Eva and {Slepian}, Zachay and {Stone}, Chris and {Stupak}, Robert and {Tarl{\'e}}, Gregory and {Li}, Ting and {Tyas}, Luke and {Vargas-Maga{\~n}a}, Mariana and {Walker}, Alistair and {Wenner}, Nicholas and {Y{\`e}che}, Christophe and {Zhang}, Yuanyuan and {Zhou}, Rongpu},
        title = "{The Spectroscopic Stage-5 Experiment}",
      journal = {arXiv e-prints},
         year = 2025,
        month = mar,
          eid = {arXiv:2503.07923},
        pages = {arXiv:2503.07923},
          doi = {10.48550/arXiv.2503.07923},
archivePrefix = {arXiv},
       eprint = {2503.07923},
 primaryClass = {astro-ph.CO},
       adsurl = {https://ui.adsabs.harvard.edu/abs/2025arXiv250307923B}
}

@ARTICLE{Reid-2016,
       author = {{Reid}, Beth and {Ho}, Shirley and {Padmanabhan}, Nikhil and {Percival}, Will J. and {Tinker}, Jeremy and {Tojeiro}, Rita and {White}, Martin and {Eisenstein}, Daniel J. and {Maraston}, Claudia and {Ross}, Ashley J. and {S{\'a}nchez}, Ariel G. and {Schlegel}, David and {Sheldon}, Erin and {Strauss}, Michael A. and {Thomas}, Daniel and {Wake}, David and {Beutler}, Florian and {Bizyaev}, Dmitry and {Bolton}, Adam S. and {Brownstein}, Joel R. and {Chuang}, Chia-Hsun and {Dawson}, Kyle and {Harding}, Paul and {Kitaura}, Francisco-Shu and {Leauthaud}, Alexie and {Masters}, Karen and {McBride}, Cameron K. and {More}, Surhud and {Olmstead}, Matthew D. and {Oravetz}, Daniel and {Nuza}, Sebasti{\'a}n E. and {Pan}, Kaike and {Parejko}, John and {Pforr}, Janine and {Prada}, Francisco and {Rodr{\'\i}guez-Torres}, Sergio and {Salazar-Albornoz}, Salvador and {Samushia}, Lado and {Schneider}, Donald P. and {Sc{\'o}ccola}, Claudia G. and {Simmons}, Audrey and {Vargas-Magana}, Mariana},
        title = "{SDSS-III Baryon Oscillation Spectroscopic Survey Data Release 12: galaxy target selection and large-scale structure catalogues}",
      journal = {\mnras},
         year = 2016,
        month = jan,
       volume = {455},
       number = {2},
        pages = {1553-1573},
          doi = {10.1093/mnras/stv2382},
archivePrefix = {arXiv},
       eprint = {1509.06529},
 primaryClass = {astro-ph.CO},
       adsurl = {https://ui.adsabs.harvard.edu/abs/2016MNRAS.455.1553R}
}

@ARTICLE{Barreira-2020,
       author = {{Barreira}, Alexandre and {Cabass}, Giovanni and {Schmidt}, Fabian and {Pillepich}, Annalisa and {Nelson}, Dylan},
        title = "{Galaxy bias and primordial non-Gaussianity: insights from galaxy formation simulations with IllustrisTNG}",
      journal = {\jcap},
         year = 2020,
        month = dec,
       volume = {2020},
       number = {12},
          eid = {013},
        pages = {013},
          doi = {10.1088/1475-7516/2020/12/013},
archivePrefix = {arXiv},
       eprint = {2006.09368},
 primaryClass = {astro-ph.CO},
       adsurl = {https://ui.adsabs.harvard.edu/abs/2020JCAP...12..013B}
}
\end{document}